\documentclass[%
 amsmath,amssymb,
aps,pre,twocolumn,amsmath,amssymb,nofootinbib,superscriptaddress,floatfix
]{revtex4-2}

\usepackage{graphicx}
\usepackage{dcolumn}
\usepackage{bm}
\usepackage{physics}
\usepackage{color}

\begin{document}

\title{Topological Floppy Modes in Epithelial Tissues}

\author{Harry Liu}
 \affiliation{Department of Physics, University of Michigan, Ann Arbor, MI 48109-1040, USA}
 
\author{Di Zhou}
\affiliation{Department of Physics, University of Michigan, Ann Arbor, MI 48109-1040, USA}
\affiliation{School of Physics, Beijing Institute of Technology, Beijing 100081, China}
\affiliation{School of Physics, Georgia Institute of Technology, Atlanta, GA 30332, USA}

\author{Leyou Zhang}
 \affiliation{Department of Physics, University of Michigan, Ann Arbor, MI 48109-1040, USA} 

\author{David K. Lubensky}
 \affiliation{Department of Physics, University of Michigan, Ann Arbor, MI 48109-1040, USA}

\author{Xiaoming Mao}
\email{maox@umich.edu}
 \affiliation{Department of Physics, University of Michigan, Ann Arbor, MI 48109-1040, USA}

\date{\today}

\begin{abstract}
Recent advances in topological mechanics have revealed unusual phenomena such as topologically protected floppy modes and states of self-stress that are exponentially localized at boundaries and interfaces of mechanical networks.  In this paper, we explore the topological mechanics of epithelial tissues, where the appearance of these boundary and interface modes could lead to localized soft or stressed spots and play a role in morphogenesis.  
We consider both a simple vertex model (VM) governed by an effective elastic energy and its generalization to an active tension network (ATN) which incorporates active adaptation of the cytoskeleton. By analyzing spatially periodic lattices at the Maxwell point of mechanical instability, we find topologically polarized phases with exponential localization of floppy modes and states of self-stress in the ATN when cells are allowed to become concave, but not in the VM.

\end{abstract}

\maketitle

\section{\label{sec:intro}Introduction}
The mechanics of epithelial tissues, where living cells closely pack a surface and mechanically interact with one another, is crucial for many morphogenetic processes, such as gastrulation, wound healing, embryogenesis, etc.~\cite{Schock2002,Sepich2012,Fristrom1988,Leptin2005,Colas2001,Trichas2012}. These processes can require particular cellular arrangements that are associated with specific mechanical properties, which have been studied intensely through analyses of the stresses and strains on the epithelial network~\cite{Cowin2007,Guillot2013,Heller2015,Lange2013,Landsberg2009,Aliee2012,Umetsu2014}. The relation between structure and mechanics in epithelial tissues not only offers a gateway for a deeper understanding of many of these natural processes but also opens possible paths to engineer potentially beneficial synthetic processes. 
In particular, boundaries and interfaces often play crucial roles in the mechanics of epithelial tissues, because they offer a platform where the dynamics of cells are most rich.
A thorough understanding of the mechanics on the boundaries and interfaces would be very helpful to characterize these phenomena.

Recent advances in theories of topological mechanics provide a fundamental framework for understanding mechanics on boundaries and interfaces of \emph{marginally stable} (i.e., ``Maxwell'') networks and how these mechanical properties are robustly controlled by topological features in the bulk~\cite{Kane2013,Lubensky2015,Mao2017}.  Many designs have been proposed utilizing topologically protected mechanical properties to produce novel cellular topological mechanical metamaterials with unusual properties such as reconfigurable surface stiffness, stress distribution, and localized modes~\cite{Paulose2015,Paulose2015a,Rocklin2017,Zhang2018}.  

Interestingly, epithelial tissues often operate at or near the verge of mechanical instability~\cite{Bi2015,Bi2016,Yan2019,Staple2010}, 
as they are then able both to support stress and to accommodate transformations during various biological processes. Moreover, it has recently been shown that topological floppy boundary modes can show up in disordered biopolymer networks when excited by active driving~\cite{Zhou2018}. 
It is thus interesting to ask whether topological mechanical properties on boundaries and interfaces can also arise in epithelial tissues.

In this paper, we study topological mechanics in epithelial tissue sheets based on two models, namely a simple (passive) vertex model (VM) and an active tension network (ATN) model~\cite{Chiou2012,Noll2017}. 
{ We adapt these models so as to put them at the Maxwell condition where the number of degrees of freedom is equal to the number of constraints.  This condition is crucial  for topological polarization to appear. 
We observe that in the ATN topologically polarized phases exist and that these phases only arise when cells become concave.
This is based on the observation that topological phase transitions in these models can only occur when edges of cells form straight lines, which leads to gap closing and only happens at the onset of convexity change.  We cannot, though, exclude the possibility of topological polarization via the creation and annhilation of Weyl points in the convex configuration.  
}

This topological polarization indicates exponentially localized floppy modes and states of self-stress on boundaries and interfaces of the system. { Mechanically,  boundaries and interfaces with exponentially localized floppy modes are much softer in comparison to other parts of the tissue.  In contrast, if a boundary does not have exponentially localized floppy modes, it would appear as rigid as the interior of the tissue.  On the other hand,  interfaces in the tissue with exponentially localized states of self-stress tend to accumulate { both external stress and internal stress} from cell activity.  
These properties are solely owing to the cell geometry of the \emph{bulk} of the tissue, instead of to special cell activities at the boundary or interface.  This is a manifestation of the ``topological protection'' of these floppy modes and states of self-stress, which endows the aforementioned phenomena with remarkable robustness: 
any weak interactions or slight changes in the geometry, as long as they do not change the topological phase of the bulk of the tissue, will not destroy the boundary and interface mechanical response. These topological mechanical properties may lead to interesting behaviors such as robust localization of dynamics or stress, and may shed new light on phenomena involving tissue boundary/interface dynamics, { such as dorsal closure \cite{Hayes2017,Kiehart2017} and invasion of larval tissue by histoblast nests \cite{Ninov2007,Ainslie2020} in \textit{Drosophila}, epiboly in teleost fish \cite{Bruce2020}, and collective migration and wound healing in a variety of \textit{in vitro} and \textit{in vivo} systems \cite{Begnaud2016,Hakim2017}.}}

\section{\label{sec:model}The Models}
In their simplest form, epithelial tissues comprise a monolayer of adjacent cells, which can often be approximated by polygons.  
Thus, the mechanics of an epithelial tissue can be analyzed by studying a two-dimensional sheet of edge-sharing polygons. A variety of variants on the basic theme of a vertex model, in which the degrees of freedom are the positions of the polygon vertices, have been proposed and can explain many observations of mechanical phenomena in epithelial tissues~\cite{Honda1983,Farhadifar2007,Fletcher2014,Bi2015,Yang2017,Merkel2018,Hufnagel2007,Salbreux2012,Spencer2017}.  
These include both passive models in which cell shapes are assumed to be governed by an (effective) energy and extensions that explicitly account for various active processes in living tissues.  Here, we reserve the term vertex model (VM) for a particular, common choice of a passive energy described below in Sec. \ref{sec:model-defns}.  We also consider an interesting example of an active model, the active tension network (ATN) model, where mechanical equilibrium is attained when both force balance at each vertex and the ``stall tension'' on each edge are reached, with the result that the tension on each edge can effectively be specified independently (subject to force balance constraints). 

In this section, we first briefly introduce the VM and the ATN to analyze the counting of the degrees of freedom and  constraints in them. We then discuss the conditions under which these models become ``Maxwell networks'', meaning that they have balanced degrees of freedom and constraints, providing the right condition for topological floppy modes to arise.  We also determine the force-balance conditions for these models, the equilibrium states of which are both stressed.

\subsection{\label{sec:model-defns}Models of epithelial cell sheets}
In the remainder of this paper, we consider tissue sheets parameterized by a set of vertex coordinates $\{\vec{R}_i\}$. We use the term VM specifically to refer to a model where the dynamics of these coordinates is assumed to be governed by a mechanical energy with the form~\cite{Honda1983}
	\begin{equation}\label{EQ:EVM}
	E = \frac{1}{2}\sum_{f} K_{P} \qty(P_f-P_{0})^2 + K_{A} \qty(A_f-A_{0})^2 ,
	\end{equation}
where $K_{P}$ is the elastic constant of cortical tension that constrains the perimeter of cells, and $K_{A}$ is an area elastic constant that could arise, for example, from an interplay between cell incompressibility in 3D and resistance to cell height differences across the tissue.  The sum is over all cells in the tissue, which are labeled by $f$ and have perimeter $P_f$ and area $A_f$.

Tissues governed by the energy of Eq.~\eqref{EQ:EVM} have been shown to exhibit a jammed phase, where any displacements of vertices cost elastic energy and the system develops a shear modulus, when the ratio  $P_0/\sqrt{A_0}$ drops below a critical value~\cite{Bi2015}.  In this jammed phase, the tissue is stabilized by an equilibrium tension, as we discuss below in the constraint counting.

In the ATN, instead of a passive tension that attempts to restore a preferred perimeter in each cell, the edges are active and try to reach a preferred ``stall tension'', determined by the local activity of the actomyosin bundle along the edge and cadherin clusters between the cells.  Mechanical equilibrium of the tissue is reached when forces balance at each vertex and each edge is at its stall tension.  Ref.~\cite{Noll2017} introduced a relaxational dynamics that specifies how the myosin concentration and the tension on each edge evolve towards this equilibrium state.  To study topological modes, however, we are only interested in small displacements from mechanical equilibrium.  In this case, we may treat the edge tensions as constants, corresponding to the long-time, elastic-like behavior of the tissue.  A similar limit was taken in Ref.~\cite{Noll2017} in the discussion of the ``isogonal'' soft modes.

Thus, for the purposes of this paper an ATN is simply a model in which each edge is endowed with a fixed tension $T_{ij}$ (where $i$ and $j$ denote the two vertices joined by the edge) 
and each cell has a pressure $\Pi_f = 2 K_A (A_f - A_0)$ conjugate to its area.  
The $T_{ij}$'s and $\Pi_f$'s must be chosen so that the net force on each vertex vanishes when the vertices are at their equilibrium positions but are otherwise arbitrary.  The model can then be viewed as having an effective energy whose differential for small vertex displacements from mechanical equilibrium is given by
	\begin{equation}\label{EQ:EATN}
	dE\qty[\{\vb{r}_i\}] = \sum_{\expval{i,j}} T_{ij}dR_{ij}+\sum_f \Pi_{f} dA_f
	\end{equation}
where $R_{ij}=\vert \vec{R}_i-\vec{R}_j \vert$ is the distance between these two vertices.

\subsection{Mechanical stability and Maxwell's counting}\label{SEC:Maxwell}
In order to analyze topological mechanics in the VM and the ATN, we need to first count the degrees of freedom and constraints in these models and identify the ``Maxwell condition'' where the balance of degrees of freedom and constraints is met. {This condition puts the system at the verge of mechanical instability, allowing unusual topologically protected modes to arise~\cite{Kane2013,Mao2017}. }

For both models, deviations from a mechanically balanced state can be described by the displacement field of the vertices $\{\vec{u}_i\}=\qty(u_1^x,u_1^y,\cdots,u_V^x,u_V^y)$ for all $V$ vertices.  Thus, each vertex $i$ displaces from its mechanical equilibrium state $\vec{r}_i$ to a new position $\vec{r}_i \to \vec{R}_i=\vec{r}_i + \vec{u}_i$.  

We now consider what constraints a displacement field in each model must satisfy to be a zero mode (ZM) that costs no elastic energy ($dE=0$).  In both the VM and the ATN, the mechanical equilibrium states we expand around are \emph{stressed}, which means edges bear nonzero tension.  As we derive in more detail in App.~\ref{APP:EE}, this results in an ``irrotational'' constraint from each stressed edge, \begin{equation} \label{ZMRot}
 (\vec{u}_i-\vec{u}_j) \cross \hat{l}_{ij} = 0 ,
	\end{equation}
where $i,j$ denote the two vertices connected by this edge, and
$\hat{l}_{ij}=(\vec{r}_{j}-\vec{r}_i)/\vert \vec{r}_{j}-\vec{r}_i\vert$ is the unit vector pointing from vertex $i$ to $j$ in the equilibrium state.

The area term of each cell contributes a constraint that the cell area needs to be preserved by any ZM (for details see App.~\ref{APP:EE}).  To set up the notation, we consider a cell with $V_f$ vertices labeled as $i=1,\ldots, V_f$, and $\vec{\mathcal{U}}_i = \vec{u}_{i+1}-\vec{u}_i$ being the relative displacement between the neighboring sites. We use $\vec{\mathcal{L}}_i = \vec{r}_{i+1}-\vec{r}_i$ to denote the vector connecting the two vertices in the equilibrium state we expand around.  The constraint that the area is preserved can then be written as
	\begin{equation} \label{ZMArea}
	\sum_{i=1}^{V_f-1}\sum_{j>i}^{V_f-1} \qty(\vec{\mathcal{U}}_j\cross \vec{\mathcal{L}}_{i} - \vec{\mathcal{U}}_i\cross \vec{\mathcal{L}}_{j}) =0 .
	\end{equation}
	
These two constraints [Eqs.~(\ref{ZMRot}, \ref{ZMArea})] are the same between the VM and the ATN.  The VM has an additional term which preserves the perimeter of each cell,
	\begin{equation} \label{ZMPeri}
	\sum_i^{V_f} \vec{\mathcal{U}}_i\cdot \hat{l}_{i,i+1} = 0 .
	\end{equation}

Therefore, the number of constraints in the VM is $N_{C}=2F+E$, where $F$ is the number of cells and $E$ is the number of edges in the network. This  follows from the fact that each cell provides a constant perimeter and a constant area constraint, and each edge provides a no-rotation constraint because it's stressed.  
In contrast, in the ATN the number of constraints is $N_{C}=F+E$, as the cell perimeter does not need to be conserved for ZMs.

The number of degrees of freedom is $N_{DOF}=2V$ in both models, because two coordinates are required to specify the position of each vertex in two dimensions.  Assuming that all vertices have coordination number $z=3$ (3 edges meet at each vertex, which is natural for polygonal tilings), we have $E=zV/2=3V/2$.  Using Euler's characteristics we have $F=E-V=V/2$.  Therefore, the total number of constraints on the VM including the area constraint is $N_{C}=5V/2>N_{DOF}$ and the model is over-constrained in the presence of stress.  The numbers of constraints and of degrees of freedom become equal when the area constraint is neglected, leading to $N_{C}=2V=N_{DOF}$, making the system a Maxwell network.

On the other hand, the ATN is a Maxwell network with the area constraint included, as in this case,  $N_{C}=F+E=2V=N_{DOF}$.

{
As we mentioned above, our choice of the elastic energy terms for these models is guided by the requirement of placing the models at the Maxwell point, { so that topological modes are permitted.}  
Thus, for the VM  we henceforth consider only the limit that the cortical tension of the cells is the dominant contribution to the energy, i.e. $K_A \rightarrow 0$.  That is, we drop the area elasticity contribution, or, equivalently, assume that pressure differences between cells are negligible. This limit has been considered in several previous studies on this model~\cite{Bi2015,Bi2016,Yan2019,Noll2017}. In contrast, for the ATN, we consider the generic case where pressure differences between cells may be significant and the area contribution cannot be ignored. 

{We emphasize that all of these choices are made solely in order to place the two models at the Maxwell point, where topologically protected zero modes are possible.  In particular, we do not claim that these particular limits have any special biophysical importance \textit{a priori}; rather, we predict that in the correct limits the VM and ATN have unusual topological properties that might be of biological interest.} 
Importantly, Refs.~\cite{Stenull2019,Sun2020,Saremi2020} show that small deviations from these ideal limits through the inclusion of weak additional terms in the energy (e.g., adding back the area term in the VM or introducing deviations from fixed tensions in the ATN) preserve the topological polarization, and only weakly lift the energy of the ZMs.  {Thus, we expect that our qualitative conclusions will continue to hold in the vicinity of the Maxwell point.}

The counting argument we give here for the VM is consistent with that provided by Bi and Yan ~\cite{Yan2019}. Their count $N_{C}=(E-E_0)+F$, where $E_0$ is the number of edges without tension, includes contributions from each tensioned edge and from each face. In our case, the energy expansion is done around a pre-stressed network where every edge carries a tension, so that $E_0 = 0$, our constraint count $N_{C}=E+F$ then matches that in \cite{Yan2019}.}

The Maxwell-Calladine index theorem asserts that in a mechanical network, the difference between the number of ZMs and the number of states of self-stress (SSSs, i.e., eigenmodes of the stress distribution leaving all components of a network in force balance) is given by the difference between the numbers of degrees of freedom and of constraints~\cite{Calladine1978,Kane2013,Lubensky2015},
\begin{align}\label{EQ:MCIT}
    N_{ZM}-N_{SSS}=N_{DOF}-N_{C} .
\end{align}
Thus, if a network is Maxwell (defined as $N_{DOF}=N_C$ in the bulk, i.e. neglecting any boundary effects), it must have equal numbers of ZMs and SSSs in the bulk. This condition means that an infinite Maxwell lattice has no ZMs unless there are SSSs. For a finite sized system under open boundary conditions, however, a subextensive number of ZMs arise due to the removed constraints on the boundary.  Whether these ZMs are localized or extensive, and where they localize, is a topologically protected property, characterized by a topological winding number~\cite{Kane2013}.  The topological state, in turn, is determined by the architecture of the tissue network, i.e., by the angles and lengths of the edges.

It is worth emphasizing that these counting arguments are done in a stressed equilibrium state in both models.  
This is different from most current models of topological mechanics.  
If there were no stress, the no-rotation constraints associated with the edges would be lifted, and the tissue would be under-constrained, with an extensive number of floppy modes.

\subsection{\label{subsec::FB}Force-balance condition}
The fact that the force equilibrium states in both the VM and the ATN are stressed requires that any choice of the state we choose to expand around, and study topological modes, needs to satisfy force balance.

\begin{figure}[h]
\includegraphics[width=0.4\textwidth]{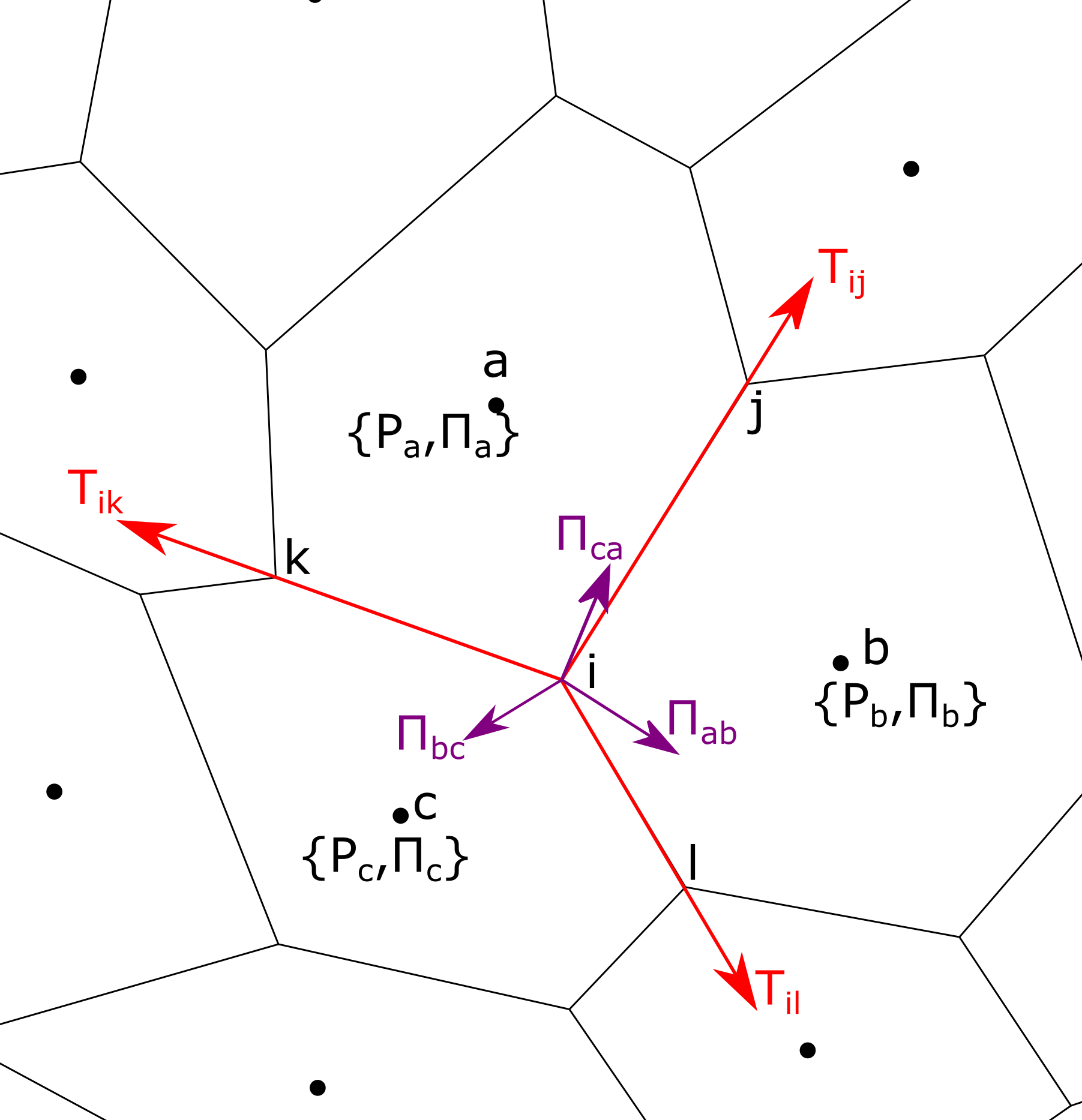}
\caption{A schematic of the variables used in the force-balance condition Eq.~(\ref{EQ:FBVM},\ref{EQ:FBATN}). The forces on vertex $i$ caused by the tension on the edges are shown as red arrows, and the forces due to the pressure of the cells are shown as purple arrows.}
\label{FIG:FB}
\end{figure}

The force balance condition can be derived by requiring $O(\vec{u})$ terms in the elastic energy to vanish [Eq.~\eqref{EQ:EVM} for the VM and Eq.~\eqref{EQ:EATN} for the ATN].  As we derive in detail in App.~\ref{APP:EE}, for the VM, after dropping area terms, this condition takes the form
\begin{equation}\label{EQ:FBVM}
	T_{ij}\hat{l}_{ij}+T_{ik}\hat{l}_{ik}+T_{il}\hat{l}_{il} = 0 ,
\end{equation}
for each site $i$, 
where as we defined above, $\hat{l}_{ij},\hat{l}_{ik},\hat{l}_{il}$ are the edge directions from vertex $i$ to vertices $j,k,l$, which are the nearest neighbors of $i$. In addition, 
\begin{align}\label{EQ:Tabc}
    T_{ij} = T_a+T_b, \quad
    T_{ik} = T_a+T_c, \quad
    T_{il} = T_b+T_c, 
\end{align}
are the tensions on the edges $ij,ik,il$ respectively, originating from the cortical tension $T_a, T_b, T_c$ from the cells $a,b,c$. The cortical tension of a cell $f$ can be calculated as
    \begin{equation}
    T_{f} = \pdv{E_f}{P_f}
    \end{equation}
for the VM.

In the ATN, as we discussed above, the Maxwell condition is satisfied when the area term is included, so the force balance condition on vertex $i$ is given by
	\begin{equation} \label{EQ:FBATN}
	\begin{split}
	& T_{ij}\hat{l}_{ij}+T_{ik}\hat{l}_{ik}+T_{il}\hat{l}_{il}\\
	& + \frac{1}{2}\Pi_{ab}\hat{n}_{ab} l_{ij}  + \frac{1}{2}\Pi_{bc}\hat{n}_{bc} l_{il}+ \frac{1}{2}\Pi_{ca}\hat{n}_{ca} l_{ik}= 0
	\end{split}
	\end{equation}
where $\Pi_{ab} = \Pi_{a}-\Pi_{b}$, $\Pi_{bc} = \Pi_{b}-\Pi_{c}$ , $\Pi_{ca} = \Pi_{c}-\Pi_{a}$ are the differences of pressures $\Pi_{a}$, $\Pi_{b}$, $\Pi_{c}$ of cells $a,b,c$ respectively.  $\hat{n}_{ab}$, $\hat{n}_{ca}$, $\hat{n}_{bc}$ are the unit vectors normal to edge $ij$ of length $l_{ij}$ pointing from cell $a$ to cell $b$, edge $ik$ of length $l_{ik}$ pointing from cell $c$ to cell $a$, and edge $il$ of length $l_{il}$ pointing from cell $b$ to cell $c$ respectively, as shown in Fig.~\ref{FIG:FB}.  The 3 terms in the second row of Eq.~\eqref{EQ:FBATN} represent the force on the vertex that comes from the pressure difference of the 3 adjacent cells. 

It is  worth noting that the tensions $T_{ij},T_{ik},T_{il}$ in the ATN are independent variables for each edge, unlike the tensions in the VM which are related to one another via Eq.~\eqref{EQ:Tabc}.

To summarize, the two main differences between the VM and the ATN, regarding mechanics around an equilibrium state, are that (i) the perimeter does not need to remain constant for  ZMs in the ATN---the edges adjust to their preferred tensions instead of returning to the preferred perimeter, and thus the area term is included in order to bring the model to the Maxwell condition, and (ii) tensions on edges in the ATN are independent on each edge, rather than determined by cortical tensions which are variables associated with cells.  As we can see in the next section, we find this condition important in allowing the system to become topologically polarized, in the cases we studied.

\section{\label{sec:topo}Topological mechanics}
In this section we investigate topological mechanics in the VM and the ATN and discuss a phase diagram of the ATN showing where topologically polarized phases arise as a function of the architecture of the cell sheet.

\subsection{Compatibility and equilibrium matrices}
The compatibility ($\vb{C}$) and equilibrium ($\vb{Q}$) matrices are the starting point to describe topological mechanics in Maxwell networks.  In simple ball-and-spring networks, these matrices map between degrees of freedom space and constraints space, and their null spaces give ZMs and SSSs, respectively.

For an epithelial cell sheet, the compatibility and equilibrium matrices need to be generalized to describe the constraints that are more complicated in nature compared to simple ball-and-spring networks.  

For the VM, as we discussed above, the constraints come from the no-rotation condition of each edge and the perimeter of each cell, so the matrix is given by
	\begin{equation}\label{CVM}
	\vb{C} \cdot u = \mqty(e^\perp\\ \Delta P)
	\end{equation}
whereas the mapping by the $\vb{Q}$ matrix is such that
	\begin{equation}
	\vb{Q} \cdot \mqty(t^\perp\\ T_P) = f
	\end{equation}
where $e^\perp$ and $t^\perp$ are $E$-dimensional vectors of transverse motion (i.e., rotation) of and force on all the edges, $\Delta P$ and $T_p$ are $F$-dimensional vectors of the changes of perimeters and cortical tensions of all the cells, $u$ and $f$ are $2V$-dimensional vectors of the displacements of and forces on the vertices. Because $F+E=2V$, both $\vb{C}$ and $\vb{Q}$ are $2V\times 2V$ dimensional square matrices.

It may appear confusing to see transverse forces $t^\perp$ on edges, whereas in the elastic energy the edges just bear cortical tension.  A way to understand it is that we are expanding around a stressed state, where the edges already carry an equilibrium stress $T$.  Displacements of vertices cause rotation of edges $e^\perp$ which leads to a \emph{transverse change} to the tensions, which is $t^\perp$.  This change of tension is perpendicular to the \emph{edge direction in the reference state}, making it $t^\perp$, but the total tension is along the displaced edge direction.  
Thus, $t^\perp$ is allowed to exist as a state of self-stress.

In the ATN, the $\vb{C}$ and $\vb{Q}$ matrices are similar to that of the VM, except that the cortical tension is replaced by an area constraint,
	\begin{equation} \label{CATN}
	\vb{C} \cdot u = \mqty(e^\perp\\ \Delta A) ,
	\end{equation}
	\begin{equation}
	\vb{Q} \cdot \mqty(t^\perp\\ \Pi) = f ,
	\end{equation}		
where $\Delta A$ and $\Pi$ are $F$ dimensional vectors of changes of area and pressure of all the cells.

In both models, similar to the ball-and-spring network models, we have $\vb{C}=\vb{Q}^T$. In the VM, because the elastic energy is conserved, a dynamical matrix can be defined as 
\begin{align}
    \vb{D} = \vb{Q} \vb{C} ,
\end{align}
which gives the quadratic expansion of energy around the equilibrium state
\begin{align}
    E = \frac{1}{2} u\cdot \vb{D} \cdot u.
\end{align}
This conserved elastic energy is not required for our discussions of topological modes.

The null-space of the $\vb{C}$ matrix and the $\vb{Q}$ matrix give ZMs and SSSs, similar to what happens in regular spring-and-mass networks. In particular, ZMs in the VM are vertex displacements that cause no rotation for the edges and no change in the perimeter of the cells, whereas ZMs in the ATN are vertex displacements that cause no rotation for the edges and no change in the area for the cells.  On the other hand, SSSs in the VM are eigenmodes of transverse forces on edges and cortical tensions on cells that leave no net force on any vertices, whereas SSSs in the ATN in this model are eigenmodes of transverse forces on edges and pressure on cells that leave no net force on any vertices.

It is worth noting that these matrices are determined by the $O(u^2)$ terms in the expansion of the elastic energy, as we discussed in Sec.~\ref{SEC:Maxwell} and App.~\ref{APP:EE}.  The $O(u)$ terms vanish when we expand around an equilibrium reference state, and lead to the force-balance condition, as discussed in Sec.~\ref{subsec::FB}.  These $O(u)$ terms do not affect the topological mechanics of the sheet.  Instead, they determine what type of reference states are allowed.
	
\subsection{Periodic epithelial sheets and topological polarization}
To explore topological mechanics in epithelial sheets, we first start from periodic lattices, for convenient analysis of topological states in momentum space.  Specifically, we consider the network topology of the tissue to be a honeycomb lattice, i.e., each cell has 6 edges and 3 edges meet at each vertex.  Real epithelial tissue can vary both in terms of the number of edges per cell, and the number of edges meeting at a vertex, but we start from this simple model for our analysis of topological mechanics.  
In particular, 
we allow the shape of the cells to deviate from a regular hexagon to tune the geometry of the network and introduce topological phases.

Specifically, we focus on the case where each unit cell of the periodic lattice contains two epithelial cells, because inversion symmetry is always preserved if we only have one epithelial cell in the unit cell, and the tissue then cannot have a topologically polarized phase ~\cite{Kane2013}.

\begin{figure}[h]
\centering
\includegraphics[width = 0.4\textwidth]{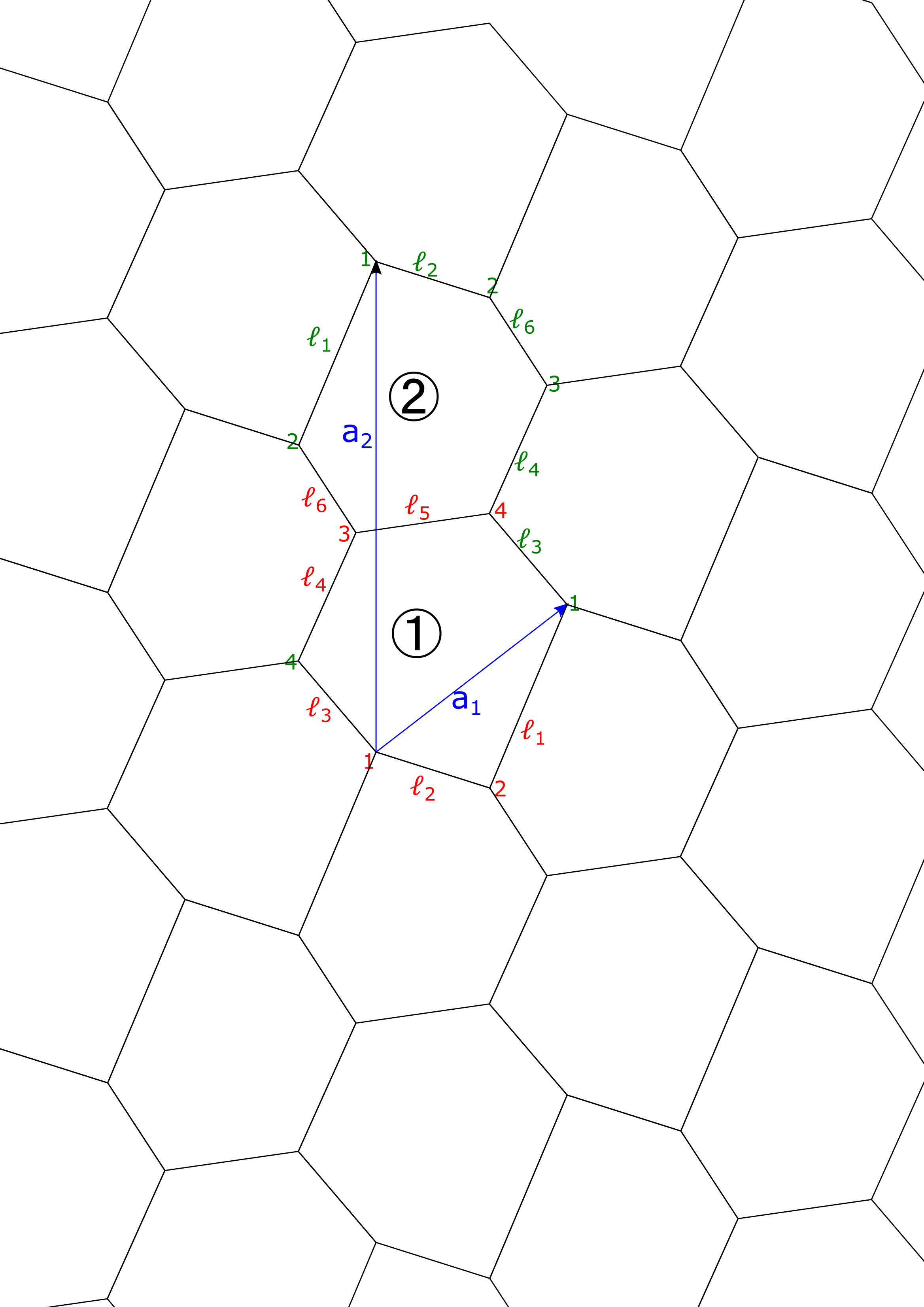}
\caption{An epithelial tissue sheet taking a periodic lattice structure with two cells $\textcircled{1}$ and $\textcircled{2}$ per unit cell.  
The basis contains 4 vertices and 6 edges as labeled in red.  The primitive  vectors $\vec{a_i}$ are labeled in blue. The same vertices and edges that are translated according to the primitive  vectors are labeled in green.
}
\label{Unit_Cell}
\end{figure}

In this 2-cell unit cell, we have 4 vertices and 6 edges in the basis, and the network is  constructed following an oblique Bravais lattice with primitive vectors $\vec{a}_1, \vec{a}_2$.
The number of degree of freedom per unit cell $n_{DOF} = 8$, and the number of constraints per unit cell $n_{C} = 6+2 = n_{DOF}$ where the 6 represents the 6 no-rotation constraints from the 6 edges, and 2 represents the  constraints associated with the two faces (perimeter for the VM and area for the ATN).

To construct the compatibility matrix, we start from the ZM conditions discussed in Sec.~\ref{SEC:Maxwell}, namely, Eqs.~(\ref{ZMRot}, \ref{ZMPeri}) for the VM, and Eqs.~(\ref{ZMRot}, \ref{ZMArea}) for the ATN.  Using these conditions, we can construct compatibility matrices $C(\vb{q})$ in momentum space that satisfy the mapping described in Eq.~\eqref{CVM} for the VM and Eq.~\eqref{CATN} for the ATN.  The null space of these compatibility matrices give the ZMs of these models.  The details of the compatibility matrix are given in the App.~\ref{APP:CM}.

The topological polarization can then be determined from these compatibility matrices, via the calculation of the winding numbers of $\det C(\vb{q})$ around the first Brillouin zone~\cite{Kane2013},
	\begin{equation}\label{EQ:WN}
	\mathcal{N}_i = \frac{1}{2\pi}\oint_{C_i} d\vb{q}\cdot \nabla_{\vb{q}} \Im \ln \det C(\vb{q}) ,
	\end{equation}
where the two paths $C_1, C_2$  wrap the first Brillouin zone along the two reciprocal vectors $\vec{b}_1, \vec{b}_2$.  A topological polarization can then be defined
\begin{equation}\label{EQ:RT}
	\vec{R}_T = -\sum_i n_i \vec{a}_i ,
\end{equation}
where $\vec{a}_i$ are the 2 primitive vectors.  Here the two integers $(n_1,n_2)$ are related to the two winding numbers calculated above by a constant shift, $n_i = \mathcal{N}_i +\Delta_i$, that accounts for the asymmetry of the choice of the unit cell, such that $\vec{R}_T$ provides a symmetric description of the polarization.  For the choice of unit cell we use, as described in Fig.~\ref{Unit_Cell},
$(\Delta_1,\Delta_2)=(2,-1)$.

\subsection{Critical configurations} 
We start our analysis of topological phases in these lattices by identifying critical configurations where ZMs are bulk modes.  These critical configurations are analogous to the regular square  and kagome lattices~\cite{Mao2017}, and the Mikado model with straight fibers~\cite{Zhou2018}, where ZMs (other than trivial translations) arise under periodic boundary conditions (PBC).

These critical configurations are vital points to construct a phase diagram for topological boundary modes in these problems. 
This can be seen from the Maxwell-Calladine index theorem [Eq.~\eqref{EQ:MCIT}].  Under PBC, Maxwell system have $N_{DOF}=N_{C}$ so in general there are no ZMs or SSSs except for the trivial translations.  
When lattices are at geometric singularities (i.e. critical configurations), such as bonds forming straight lines, additional SSSs arise under PBC, leading to additional ZMs, because $N_{ZM}-N_{SSS}=0$ is always satisfied.  These ZMs and SSSs are \emph{bulk modes} as opposed to boundary modes, as they satisfy PBC.

\begin{figure}[h!]
	\centering
    \includegraphics[scale=0.5]{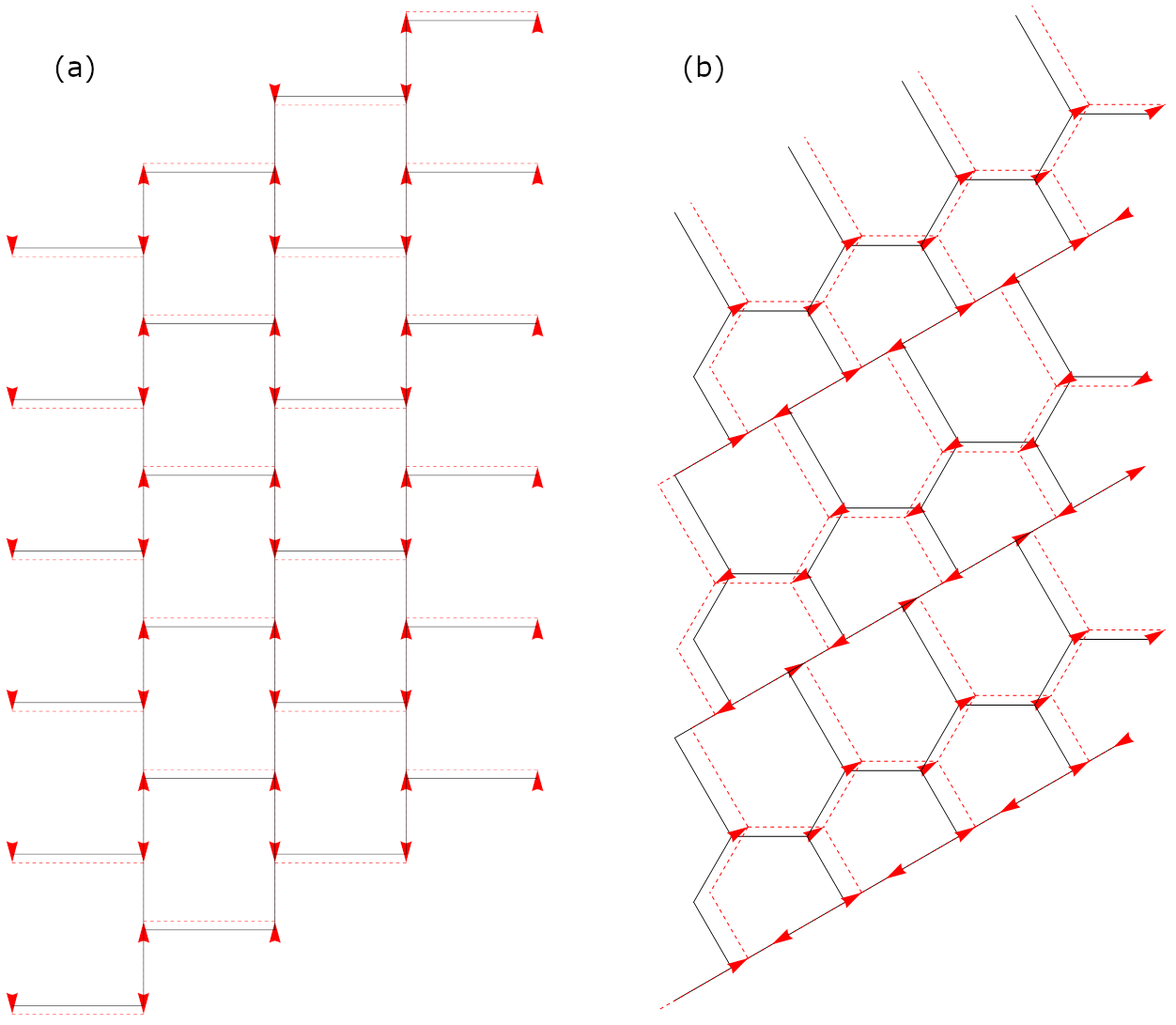}
    \caption{Critical configurations with edges of cells forming straight lines along $\vec{a}_2$ (a) and $\vec{a}_1$ (b).  Examples of bulk ZMs in these configurations are shown with red arrows (vertex displacements) and dashed lines (deformed configurations).}
\label{bm}
\end{figure}

Similarly, the VM and the ATN also 
develop these SSS-ZM pairs under PBC when edges of the cells form straight lines.  There are two such critical configurations in the 2-cell unit cell lattice, as shown in Fig.~\ref{bm}, and they give rise to bulk ZMs.  In these critical configurations,  edges in these straight lines can carry equal $t^{\perp}$ which are balanced on all nodes, giving rise to SSSs.  The corresponding ZMs are shown in Fig.~\ref{bm}, where cells in each straight vertical ``strip'' shift relative to one another, leaving all edges parallel to their original direction.  These ZMs preserve all edge directions, perimeter, and area,  so they are ZMs in both the VM and the ATN.
Our lattices yield these two types of critical configuration because of our choice of the 2-cell unit cells.  Other critical configurations involve wider strips can also arise when one chooses bigger unit cells.

At these critical configurations, the lowest phonon band has $\omega=0$ lines due to these bulk ZMs, and the momentum space $\omega=0$ lines are perpendicular to these straight lines in the real-space lattice.  As a result, the topological winding number [Eq.~\eqref{EQ:WN}] is ill-defined at these critical configurations.  
The system becomes gapped when the geometry is perturbed, leading to phases with different topological polarizations, as we discuss below.

\subsection{Polarized Phases}
In order to search for topologically polarized phases, we choose to study geometries of cell sheets perturbed around the critical configuration in Fig.~\ref{bm}(a) where straight lines of edges form along the $\vec{a}_2$ direction, as it is a simple geometry with high symmetries.  In this analysis we find topologically polarized phases in the ATN, which we discuss below.  Due to fewer free parameters in choosing force-balanced reference states, the VM does not show any topologically polarized phases.  We will comment on this at the end of this section. 

\begin{figure}[h!]
	\begin{center}
    \includegraphics[scale=0.5]{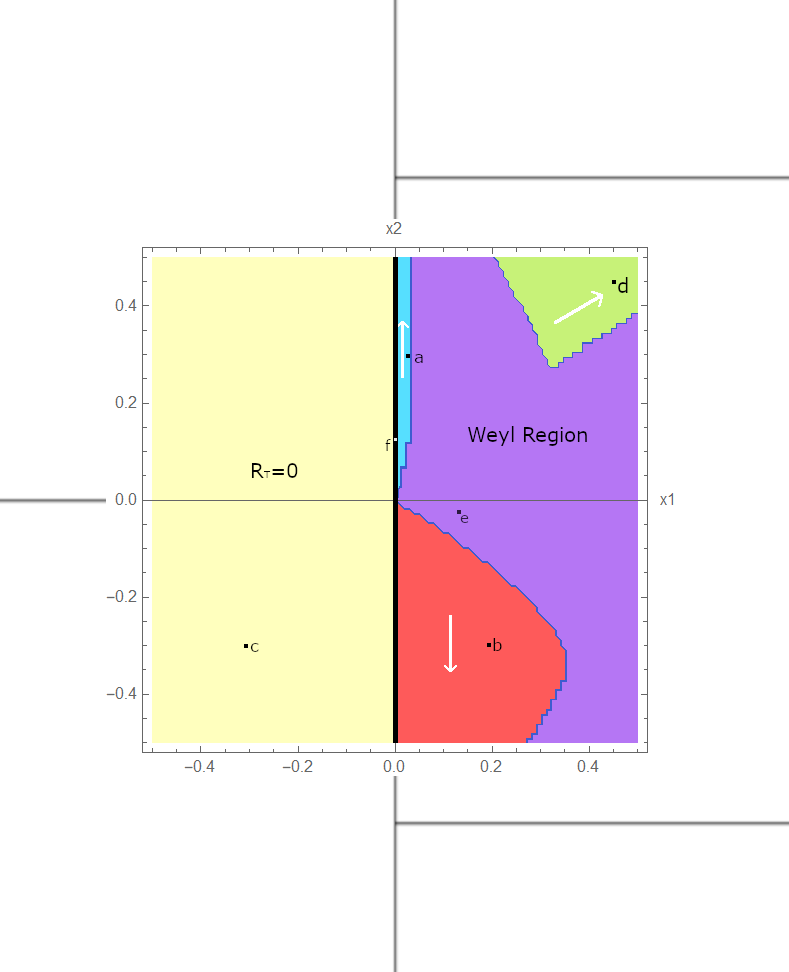}
    \end{center}
\caption{
Topological phase diagram of the cell sheet lattice as an ATN around critical configuration in Fig.~\ref{bm}(a).  The geometry of the lattice is such that vertices $1,3,4$ stay fixed, while vertex 2 is displaced by $(x_1,x_2)$ which are the axes of the phase diagram.
The phase diagram is overlaid on the real space lattice to make the geometry  clear.  
The thick black line marks  critical  configurations, and 5 different topological phases are observed.  The yellow region is un-polarized.  
The cyan, red, and green regions are topologically polarized with $\vec{R}_T$ along $\vec{a}_2$, $-\vec{a}_2$, and $\vec{a}_1$ respectively, as indicated by the white arrows.  In the purple region the lattice displays Weyl points and thus topologically protected bulk floppy modes.  Six representative configurations of these regions (marked by black dots) are shown in Fig.~\ref{FIG: Example_Network}.
}
\label{pd2}
\end{figure}

To construct a phase diagram for the ATN, we place vertex number 2 (as labeled in Fig.~\ref{Unit_Cell}) at different positions, which breaks the straight lines and lift the bulk ZM-SSS pairs.  At each given displacement $(x_1,x_2)$ of vertex number 2 [from the critical state Fig.~\ref{bm}(a)], we define a new lattice (which is a distinct reference state), and calculate winding numbers using Eq.~\eqref{EQ:WN}.  The result is shown in Fig.~\ref{pd2}.  
The phase diagram around critical configuration in Fig.~\ref{bm}(b) is included in the App.~\ref{APP:Pa1}.

\begin{figure*}[t]
	\centering
    \includegraphics[scale=0.28]{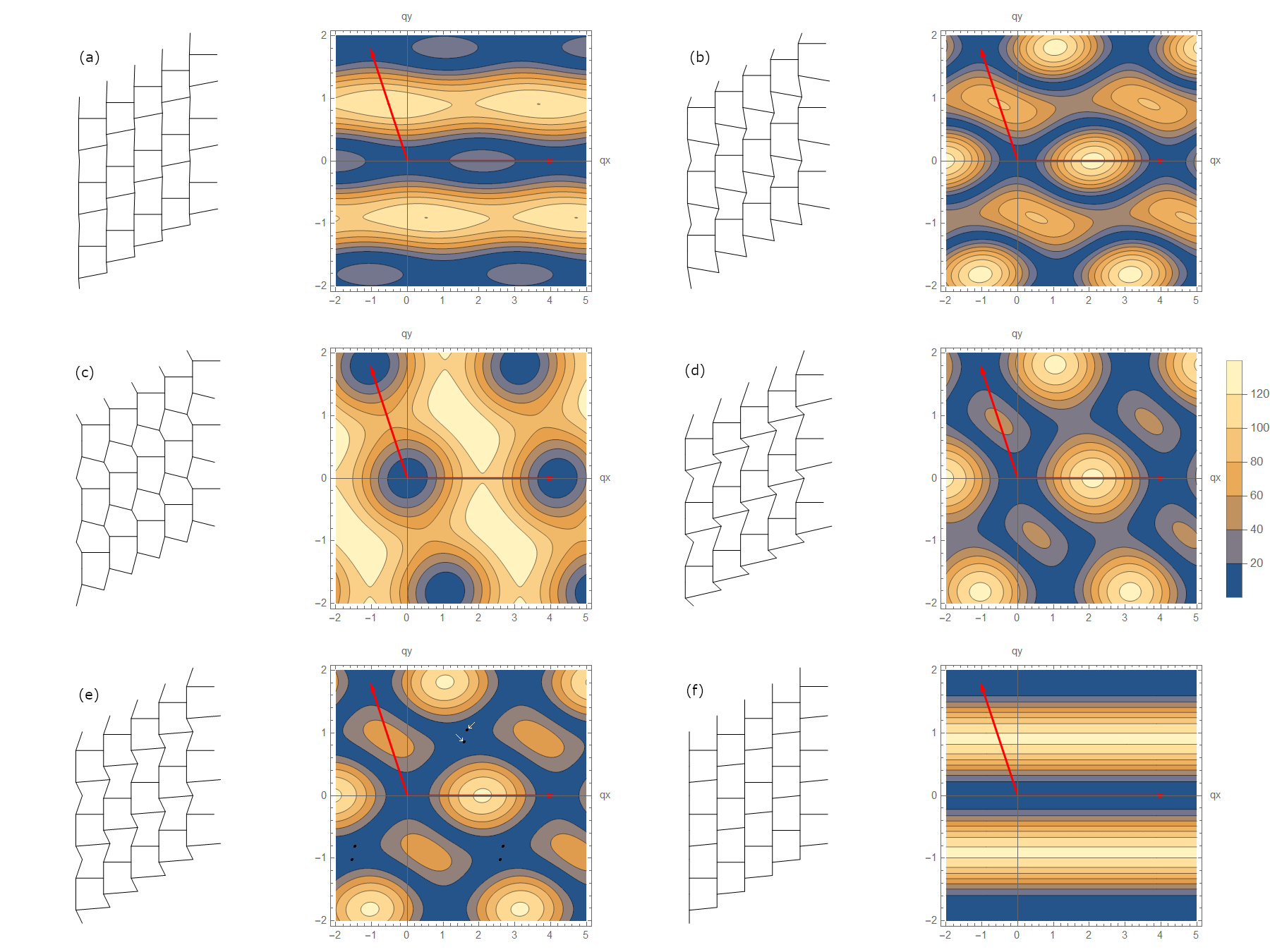}
    \caption{Representative examples of cell sheet lattices in different regions of the phase diagram (Fig.\ref{pd2}).  (a) A polarized lattice with $\vec{R}_T = \vec{a}_2$. (b) A polarized lattice with $\vec{R}_T = -\vec{a}_2$.  (c) An unpolarized lattice. (d) A polarized lattice with $\vec{R}_T = -\vec{a}_1$. (e) A lattice with Weyl modes. (f) A lattice at critical configuration.  For each panel, the real space lattice is shown on the left and the phonon dispersion relations ($\omega$  as a function of $q_x,q_y$) is shown on the right.  The two red arrows show the reciprocal lattice vectors $\vec{b}_1,\vec{b}_2$. Note the Weyl  points in (e) represented as black dots.}
\label{FIG: Example_Network}
\end{figure*}

A few interesting features arise in this phase diagram.  First, as vertex number 2 moves vertically along the straight lines, the system stays critical, as the bulk modes of shifting cells vertically remain being ZMs.  Second, as vertex number 2 moves to the left, all cells become convex, and the cell sheet is always \emph{unpolarized} ($\vec{R}_T=0$) in this type of geometry.  Third, as vertex number 2 moves to the right, all cells become concave, and the  sheet can become polarized up or down, separated by a region where Weyl modes arise.  

\begin{figure}[h!]
	\centering
    \includegraphics[scale=0.158]{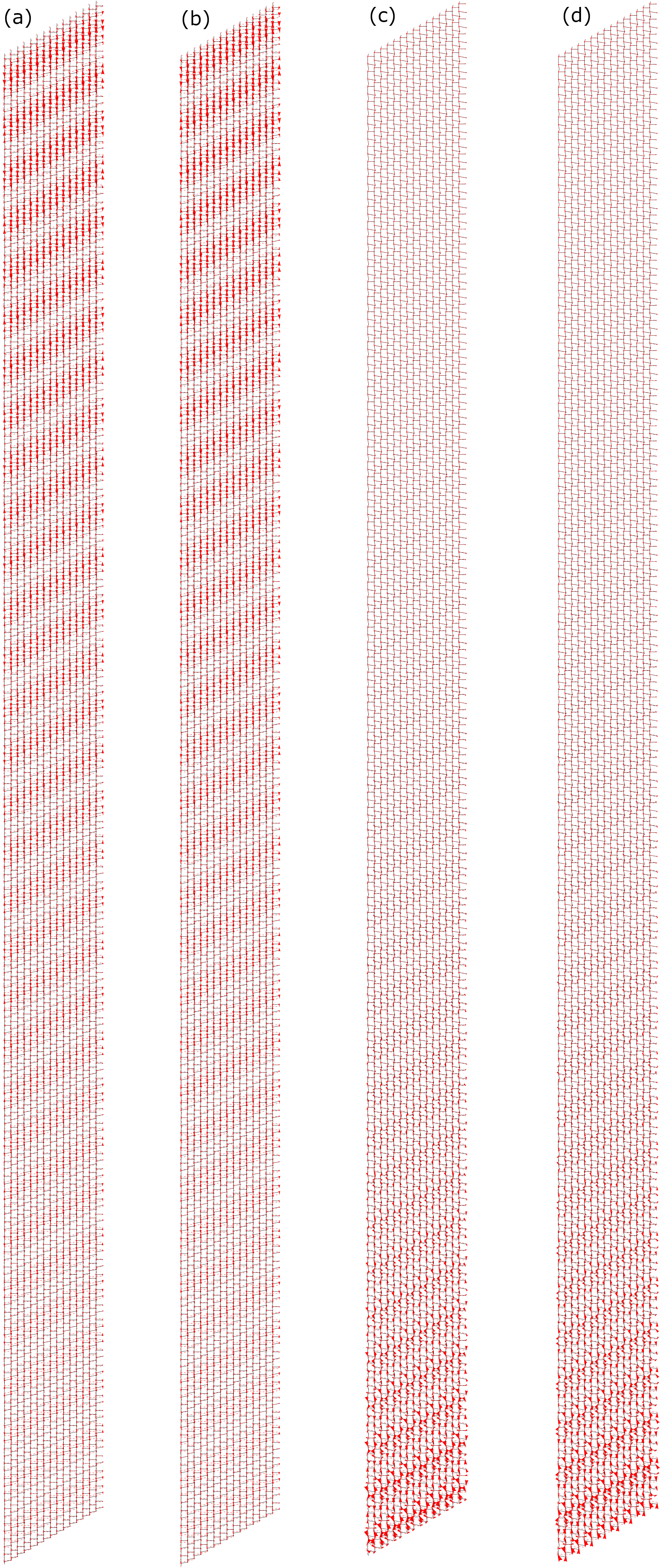}
    \caption{Examples of topological ZMs in polarized ATNs. (a-b) A lattice with $\vec{R}_T = \vec{a}_2$ [same as the lattice in Fig.~\ref{FIG: Example_Network}(a)] shows two ZMs both localized on the top boundary. (c-d) A lattice with $\vec{R}_T = -\vec{a}_2$ [same as the lattice in Fig.~\ref{FIG: Example_Network}(b)] shows two ZMs both localized on the bottom boundary. The ZMs are calculated with PBC along the $\vec{a}_1$ direction, taking a wavevector $\vec{q}$ such that $\vec{q}\cdot \vec{a}_1 = \pi$.
    }
\label{topo_zm}
\end{figure}

Some representative configurations of these phases and their phonon dispersion relations are shown in Fig.~\ref{FIG: Example_Network}.  We also plot some topological boundary ZMs for configurations with $\vec{R}_T = \pm\vec{a}_2$ [(a,b) in Fig.~\ref{FIG: Example_Network}] in Fig.~\ref{topo_zm}, where the ZMs are localized at the top and bottom edges respectively.  To make these plots, we take fixed wave numbers along the lattice boundary parallel to $\vec{a}_1$ and have PBC along this direction.  We have open boundary conditions at the top and bottom boundaries and calculate these ZMs.  It is visible from these plots that the modes preserve the edge directions and the cell areas, and are indeed ZMs of the sheet.  

For each column of unit cells, two ZMs emerge due to the open boundary on the top and the bottom.  This can be seen from Fig.~\ref{Unit_Cell} where cutting an open boundary along $\vec{a}_1$ removes two constraints (one edge and one area) per column of unit cells.  In the topologically polarized phases with $\vec{R}_T = \vec{a}_2$ [Fig.~\ref{FIG: Example_Network}(a)] both modes are localized on the top boundary [Fig.~\ref{topo_zm}(a,b)], leaving the bottom boundary rigid because it is ZM free.  
In the topologically polarized phases with $\vec{R}_T = -\vec{a}_2$ [Fig.~\ref{FIG: Example_Network}(b)] both modes are localized on the bottom boundary  [Fig.~\ref{topo_zm}(c,d)]], leaving the top boundary rigid.  

It is interesting to note that the decay length of the $\vec{R}_T = \vec{a}_2$ configuration appears to be very  long [Fig.~\ref{topo_zm}(a)].  This is due to the fact that the polarized phase with $\vec{R}_T = \vec{a}_2$ is a very narrow region on the phase diagram.  Note that at the critical phase, the decay length is infinity (the ZMs are bulk modes).  
As a result, the geometric perturbation of the unit cells in this phase is not large enough to significantly decrease the decay length of the ZMs before hitting Weyl modes configurations. 

\begin{figure}[h!]
\includegraphics[width=0.3\textwidth]{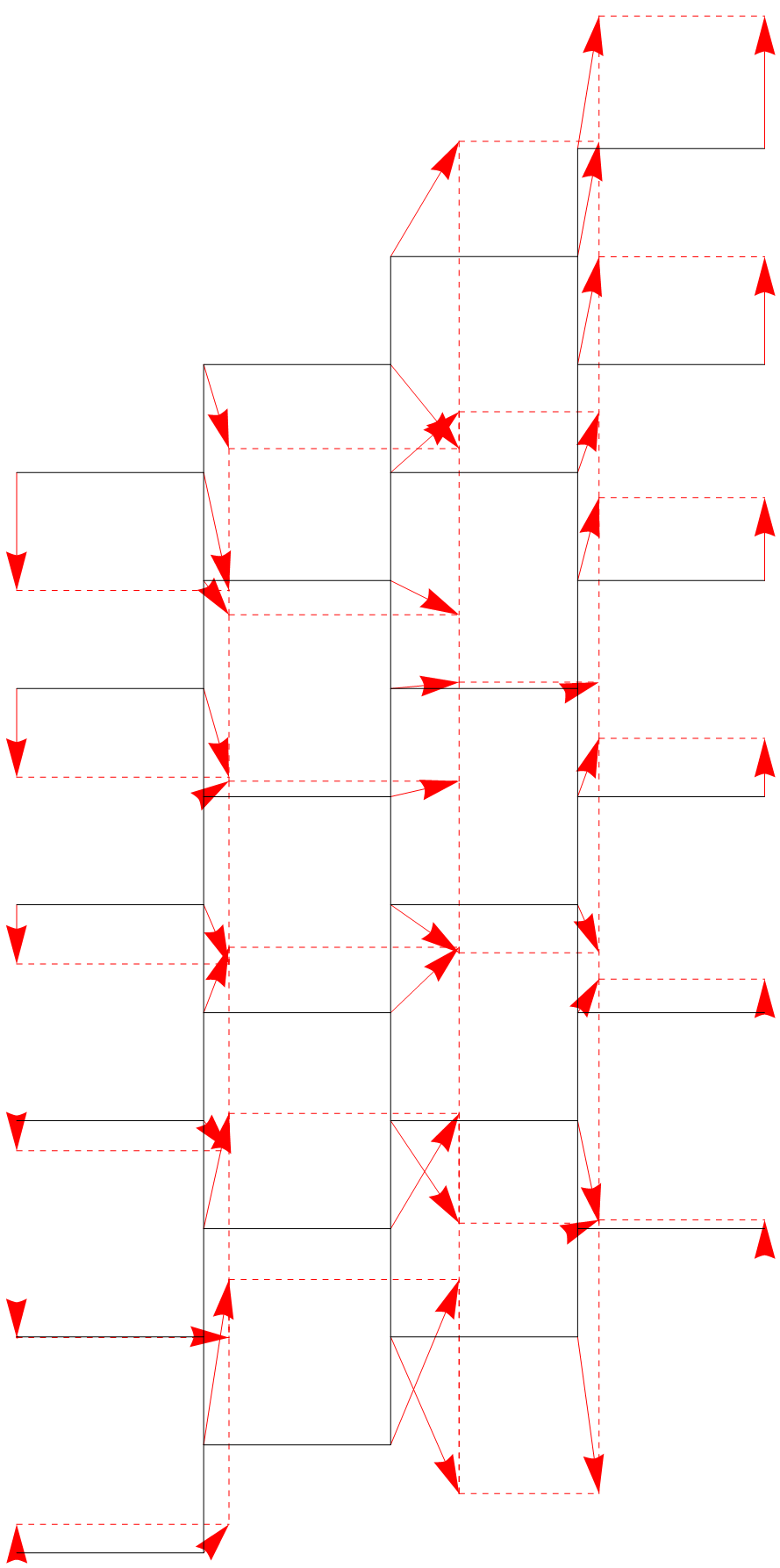}
\caption{The "breathing" mode at the critical configuration, where the straight strips get thinner and broader in an alternating order. The network is under PBC for the left-right boundary, and open boundary condition for the top-bottom boundary.}
\label{squeeze_mode}
\end{figure}

At critical configurations [Fig.~\ref{FIG: Example_Network}(f)], as we mentioned above, the ZMs are bulk modes.  For these lattices, one of the two ZMs per column is the same as the ZM computed under PBC [Fig.~\ref{bm}(a)], whereas the other one involves an interesting ``breathing'' motion of the columns of unit cells, as shown in Fig.~\ref{squeeze_mode}.  One might think of this mode as a boundary mode because of its seemingly larger magnitude of displacement on the top and bottom.  However, this is a bulk ZM, since the displacements increases linearly from the center to the boundaries (instead of exponential growth), resulting in constant strain.  The ZMs of the topologically  polarized phases can be seen as the evolution and linear combination of these two bulk ZMs at the critical configuration.

All these configurations can satisfy force balance in the ATN, by properly choosing tension of the edges and pressure of the cells.  This can be seen by considering these configurations as mechanical networks with central force springs and pressure on cells (as discussed in Sec.~\ref{subsec::FB}).  Given the hexagonal topology  of the cells and all vertices at $z=3$, the network is Maxwell regarding pre-stress, so there must be at least two global SSSs that make the system force balance under PBC at any geometry~\cite{Mao2017}.  It is worth pointing out that, by definition, all models in which the only degrees of freedom are the vertices ignore the curvature of the cell edges induced by the pressure difference between the cells, which corresponds to taking the bending stiffness of the edges to be large.  In addition, as we mentioned above, topological polarization in these cell sheet lattices requires concave cell shapes. { In this situation, force balance at the vertex with the concave angle might typically be expected to require active compression (i.e. negative $T_{ij}$) on at least one edge. In particular, it is clearly the case that some $T_{ij}$ must be negative if pressure differences between cells are small enough.  Although we cannot categorically exclude that some equilibrium configuration with concave cells and large pressure differences exists where all of the tensions are positive, we also have never been able to come up with such a counterexample.  We thus hypothesize that topological polarization normally requires negative tensions.
}

Coming back to the VM, the reason that the VM doesn't show any topological polarization is due to the more constraining force-balance condition in the VM.  As shown in Eq.~\eqref{EQ:Tabc}, instead of freely chosen cell edge tensions as in the ATN, cell edge tensions in the VM come from cortical tensions of the cells, and must satisfy Eq.~\eqref{EQ:Tabc}.  Consequently, the number of free parameters of equilibrium states is reduced.  In particular, 
for the 4 sites in the unit cell, according to the force balance equation in Eq.~\eqref{EQ:FBVM} and the cortical tension equation Eq.~\eqref{EQ:Tabc}, we have 
    \begin{eqnarray}
    -\qty(T_1+T_2) \vec{l}_2 + \qty(2T_1) \vec{l}_3 + \qty(T_1+T_2) \vec{l}_1 &=& 0,\\ \label{EQ:FBsite1}
    -\qty(T_1+T_2) \vec{l}_2 + \qty(2T_2) \vec{l}_6 + \qty(T_1+T_2) \vec{l}_1 &=& 0,\\ \label{EQ:FBsite2}
    -\qty(T_1+T_2) \vec{l}_4 + \qty(2T_2) \vec{l}_6 + \qty(T_1+T_2) \vec{l}_5 &=& 0,\\ \label{EQ:FBsite3}
    -\qty(T_1+T_2) \vec{l}_4 + \qty(2T_1) \vec{l}_3 + \qty(T_1+T_2) \vec{l}_5 &=& 0,   \label{EQ:FBsite4} 
    \end{eqnarray}
where $T_1,T_2$ are the cortical tensions of the two cells in the unit cell.  
Eq.~(\ref{EQ:FBsite1}-\ref{EQ:FBsite4}) impose an additional constraint of the force equilibrium state $\qty(T_1)\vec{l}_3 = \qty(T_2)\vec{l}_6$. { This extra constraint means that $\vec{l}_3$ is parallel to $\vec{l}_6$ which makes the 2 cells in the unit cell to have almost the same geometry with only length difference for edges $\vec{l}_3$ and $\vec{l}_6$ (see Fig.~\ref{Unit_Cell}).
Thus under inversion around the center of edge $\vec{l}_5$, the unit cell overlaps itself up to only the length difference, with all cell-edge angles preserved. 
As we discussed above, the only way the unit-cell geometry enters the mechanics of the cell sheet in the VM is through these edge angles, and the lengths of the edges are irrelevant.  As a result, the mechanics of the VM with two cells per unit cell always has \emph{inversion symmetry due to force balance}, and thus cannot topologically polarize.  The only way to polarize the VM is to either allow larger unit cell (3 or more cells in each unit cell and not under columnar arrangement) or to introduce disorder.
}
Alternatively, force-balance constraints may be lifted by allowing external forces imposed by the substrate, which offers us a larger parameter space to have topologically polarized cell configurations.

The  discussions in this section are all based on periodic lattices.  They could potentially be extended to generic, disordered cell sheets with the connectivity topology of honeycomb networks.  In App. \ref{APP:TM} we sketch a transfer matrix method~\cite{Zhou2018} that can be applied to disordered cell sheets to investigate possible topological phases in future studies.  Detailed studies of disordered cell sheets are beyond the scope of this paper.

\section{\label{sec:disc}Discussion}
In this paper, we study topological mechanics in two theoretical models of epithelial tissues, namely the VM and the ATN. We identify topologically polarized phases in the ATN where ZMs and SSSs localize on boundaries of the tissue in a topologically polarized way.  
{ In contrast, the VM doesn't support topologically polarized phases in the periodic lattice structures we studied, due to the more constraining nature of its force-balance condition. { It is in principle possible that topological polarization could develop in the VM if larger unit cells or disordered configurations are considered, but such configurations are beyond the scope of the current paper.} }

In order to study topological mechanics in the ATN, we place the system at the Maxwell point with balanced degrees of freedom and constraints. {To this end, we consider a generic case where the pressure difference between cells is not ignored so the area constraint needs to be included}. 
We study these cell sheets in a periodic lattice setting of honeycomb topology, and the unit cell consists of two cells, for simplicity. Our results show that all lattices of convex cells are topologically equivalent and do not show any topological polarization.  Topologically polarized phases arise when the cells become concave, which usually implies that some cell edges carry active compression rather than tension. These topologically  polarized phases are characterized by exponentially localized ZMs and SSSs on boundaries and interfaces pointed to by the topological polarization vector. {This indicates that when non-convex cellular shapes are experimentally observed on epithelial tissues, one may expect to discover topologically-polarized mechanical phenomena as we discuss here.  
}

Mechanical topological polarization results in strongly asymmetric mechanical responses, similar to what have been observed in spring-and-mass models~\cite{Paulose2015,Paulose2015a,Rocklin2017,Mao2017,Zhou2018,Zhou2019}. Namely, if the topological polarization $\vec{R}_T$ points towards an open edge, it exhibits extra exponentially localized ZMs, while an open edge on the opposite side loses ZMs and may become rigid if all ZMs are polarized to the opposite side.  Moreover, interfaces connecting domains of tissue of opposite topological polarization can host exponentially localized ZMs (SSSs) due to the accumulation of ZMs (SSSs) directed by $\vec{R}_T$.

{
Topologically protected ZMs induce localized softness at boundaries and interfaces of the tissue.  Compared to normal tissues, where boundaries are usually softer than the bulk in an isotropic way, topologically polarized tissues exhibit softness in a highly anisotropic manner, where some boundaries (ones opposite to the direction of $\vec{R}_T$) appear to be as rigid as the bulk, and some interfaces (ones with accumulated ZMs due to different $\vec{R}_T$ from domains around them) may be as soft as a normal boundary.

Similarly, topologically protected SSSs induce unusual local stiffness.  As shown in Refs.~\cite{Paulose2015a,Zhang2018}, when a material is under external load, stress is ``attracted'' to interfaces with  localized SSSs.  Biologically, this elevated local stress may  cause interesting consequences in cells at these interfaces.
}

In addition, even in the bulk of a  topologically polarized tissue far from boundaries or interfaces, the mechanical response to local perturbations (from cell activity or from external forces) can show strong directionality.  It has been shown in Ref.~\cite{rocklin2017directional} that in a topologically polarized mechanical network  stress and displacement propagate in opposite directions.

It is worth pointing out that we made the simplifying assumption that having boundaries and interfaces does not interfere with the active stresses in the tissue sheet.  Rigorously speaking, force balance may be violated at these boundaries and interfaces.  This will lead to local deformations to re-balance the stress, causing locally perturbed geometry at the boundaries and interfaces.  Alternatively, these active stresses can be balanced by external forces from the substrate the cell sheet attaches on (the extracellular matrix) or other biological components in contact with the sheet, so that the homogeneous lattice configurations are maintained. We conjecture that the topological mechanical properties will survive despite these perturbations, given their topological robustness.  It has been recently shown that topological mechanical properties are indeed robust against various perturbations from disorder to stress~\cite{Zhou2018}, and random damage~\cite{Zhang2018} of the networks.  Detailed numerical studies of these cell sheets with actual open boundaries and interfaces will be the subject of future studies.

Biologically, these topologically robust mechanical properties may lead to interesting consequences.  When cells are arranged such that ZMs localize at certain boundaries and interfaces, the greatly  decreased local stiffness may allow significant changes of cell shape and trigger special biological processes.  On the other hand, when cells are arranged such that SSSs localize at certain interfaces, stress significantly increases at these locations, which may trigger processes such as cell proliferation or the cell sheet to buckle out-of-plane at these controlled locations.

\begin{acknowledgments}
This work is supported by the National Science Foundation (HL, DZ, LZ, and XM, Grant No. NSF-EFRI-1741618) and by a Margaret and Herman Sokol Faculty Award (DKL).

\end{acknowledgments}

\appendix
\begin{widetext}
\section{Expansion of Elastic Energy}\label{APP:EE}
In this appendix we expand the elastic energy of both the VM and the ATN, as stressed elastic media, and derive the force balance condition from the first order terms of the expansion and constraints for ZMs from the second order terms of the expansion.
\subsection{Elastic energy}
The change of the elastic energy in both the VM and the ATN can be generically written as 
    \begin{equation}
    d E = \sum_{\expval{ij}}\tilde{T}_{ij}d R_{ij} + \sum_{f}\tilde{\Pi}_f d A_f .
    \end{equation}
This expression takes the same form as the differential elastic energy of the ATN [Eq.~\eqref{EQ:EATN}], but it also applies to the VM when it is considered an expansion of Eq.~\eqref{EQ:EVM} where the $\tilde{T}_{ij}$'s come from the cortical tensions as we discuss below. 
We introduce a $2V$ dimensional vector 
$\vb{u}=\qty(u_1^x,u_1^y,\cdots,u_n^x,u_n^y)$ to denote the displacement of all vertices, and expand the elastic energy change up to the 2nd order in $\vb{u}$. We add a tilde on the tension and pressure, $\tilde{T}_{ij}, \tilde{\Pi}_f$ to denote that they may contain $\order{u}$  terms.

Between the VM and the ATN, the major difference is reflected in the edge tension $T_{ij}$.  These tensions are controlled by the cortical tensions $T_a$ and $T_b$ of the adjacent cells $a$ and $b$ in the VM, whereas in the ATN they are adjusted by the myosin dynamics on the edge to reach their stall values. Thus, as discussed in Sec.~\ref{sec:model-defns}, we assume here that the edge tensions in the ATN are constant,
\begin{equation}
  \tilde{T}_{ij}= T_{ij}
\end{equation}
for the ATN but
    \begin{equation} \label{Tij}
    \begin{split}
    \tilde{T}_{ij} &= K_{P}\left\lbrack(P_{a}-P_0) + (P_{b}-P_0) +\frac{1}{2}dP_{a}+\frac{1}{2}dP_{b}
    \right\rbrack\\
    &= T_a + T_b + \frac{K_{P}}{2}\qty(dP_{a}+dP_{b}) \;
    \end{split}
    \end{equation}
for the VM, where the cell perimeters are to be evaluated at the equilibrium configuration. This expression comes from an expansion of the cortical tension term of Eq.~\eqref{EQ:EVM} around a stressed state with ``pre-stretch'' $P_{f}-P_0$.  
The change of perimeter for each cell can be expressed in  terms of the displacement field $\vb{u}$.  To second order  we have
    \begin{equation} \label{dPf}
    dP_{f} = \vb{u} \cdot \nabla P_f + \frac{1}{2}\qty(\vb{u}\cdot \nabla \nabla^T P_{f}\cdot \vb{u}^T)
    \end{equation}
where $\nabla = \qty(\partial_1^x, \partial_1^y, \cdots, \partial_n^x, \partial_n^y)$ and $\nabla \nabla^T $ is the Hessian matrix. Here the differential is taken with respect to $\vb{u}$ so that $\partial _1 ^x = \partial /\partial u_1^x$.  

The area contributions  are treated the same in both models including the  change in pressure due to the displacements $\vb{u}$,
    \begin{equation} \label{Pif}
    \tilde{\Pi}_{f} = K_{A}\left\lbrack\ (A_f-A_0) + \frac{1}{2}dA_{f}\right\rbrack 
    = \Pi_f +\frac{K_A}{2}dA_f \;,
    \end{equation}
where $A_f$ is to be evaluated at mechanical equilibrium.  This expression comes from an expansion of the area term of Eq.~\eqref{EQ:EVM} around a stressed state with ``pre-area-expansion'' $A_{f}-A_0$. 
The change of cellular area $dA_f$ can be expanded as (to second order)
    \begin{equation} \label{dAf}
    dA_{f} = \vb{u} \cdot \nabla A_f + \frac{1}{2}\qty(\vb{u}\cdot \nabla \nabla^T A_{f}\cdot \vb{u}^T)
    \end{equation}

Thus, combining Eq. \eqref{Tij} to Eq. \eqref{dAf} we obtain the 2nd order expansion of $dE$ with respect to the displacement field $\vb{u}$ in the VM as
    \begin{equation} \label{expansionVM}
	\begin{split}
	dE_{\text{VM}} =& \sum_{\expval{ij} }\qty(T_a+T_b)\qty(\vb{u}\cdot \nabla R_{ij}) + \frac{K_{P}}{2}\qty[\qty(\vb{u}\cdot \nabla (P_{a}+P_{b}))\qty(\vb{u}\cdot \nabla R_{ij})]
	+ \frac{\qty(T_a+T_b)}{2}\qty(\vb{u}\cdot \nabla \nabla^T R_{ij} \cdot \vb{u}^T) \\
	&+ \sum_f \Pi_f \qty(\vb{u} \cdot \nabla A_f) +  \frac{K_{A}}{2}   \qty(\vb{u}\cdot \nabla A_f)^2+ \frac{1}{2}\Pi_f \qty(\vb{u}\cdot \nabla \nabla^T A_{f}\cdot \vb{u}^T) + \order{\vb{u}^3}
	\end{split}
	\end{equation}

Similarly, for the ATN, we have the energy expansion
    \begin{equation} \label{expansionAT}
    \begin{split}
    dE_{\text{ATN}} =& \sum_{\expval{ij} }T_{ij}\qty(\vb{u}\cdot \nabla R_{ij}) + \frac{T_{ij}}{2}\qty(\vb{u}\cdot \nabla \nabla^T R_{ij} \cdot \vb{u}^T) 
     \\
    &+ \sum_f \Pi_f \qty(\vb{u} \cdot \nabla A_f) + \frac{K_{A}}{2}   \qty(\vb{u}\cdot \nabla A_f)^2
    + \frac{1}{2}  \Pi_f \qty(\vb{u}\cdot \nabla \nabla^T A_{f}\cdot \vb{u}^T) + \order{\vb{u}^3}
    \end{split}
    \end{equation}

\subsection{Force-balance condition}
The force balance condition comes from the fact that the  expansions of Eq.~\eqref{expansionVM} and \eqref{expansionAT} must have vanishing $\order{u}$ terms, so that there is no net force on any vertex. This condition takes the form
    \begin{equation} \label{FB}
    dE^{(1)}=\sum_{\expval{ij}}T_{ij} \qty(\vb{u} \cdot \qty(\nabla R_{ij})) + \sum_f \Pi_f \qty(\vb{u}\cdot \nabla A_f) = 0
    \end{equation}
for any choice of the displacement field $\vb{u}$.
This equation is exactly the force balance condition described in Eq.~\eqref{EQ:FBATN} of the main text for the ATN. 
For the  VM, we keep only the first term in Eq.~\eqref{FB}, because we drop the area term in order to satisfy Maxwell's condition. This leads to Eq.\eqref{EQ:FBVM} of the main text. 

\subsection{The Hessian}
Now we turn to examine the $\order{u^2}$ terms in the expansions and identify the constraints. For the VM, as discussed in the main text, we only treat the cortical tension as dominant contribution to the elastic energy in order to place the model to the Maxwell condition.  The 2nd order terms in Eq.\eqref{expansionVM} are thus
    \begin{equation}
    \begin{split}
    dE_{\text{VM}}^{(2)} &= \sum_{\expval{ij}}\frac{K_{P}}{2}\qty[\qty(\vb{u}\cdot \nabla (P_{a}+P_{b}))\qty(\vb{u}\cdot \nabla R_{ij})]
    + \qty(T_a+T_b)\qty(\vb{u}\cdot \nabla \nabla^T R_{ij} \cdot \vb{u}^T)
	\end{split}
    \end{equation}

The $\order{u^2}$ terms for the ATN from Eq. \eqref{expansionAT} are
    \begin{equation}
    \begin{split}
    dE_{\text{ATN}}^{(2)} =& \sum_{\expval{ij}}\frac{T_{ij}}{2}\qty(\vb{u}\cdot \nabla \nabla^T R_{ij} \cdot \vb{u}^T)
    + \sum_f \frac{K_A}{2}   \qty(\vb{u}\cdot \nabla A_f)^2
    + \frac{1}{2} \Pi_f  \qty(\vb{u}\cdot \nabla \nabla^T A_{f}\cdot \vb{u}^T)
    \end{split}
    \end{equation}

These $\order{u^2}$ terms lead to an elastic energy that consists of all complete square terms.  Because these complete square terms must all be zero to make the elastic energy vanish, they provide constraints discussed Sec.~\ref{SEC:Maxwell} in the main text.  We derive these complete square terms below.  

For a cell $f$ modeled as an $V_f$-polygon, the perimeter and area can be written as
	\begin{eqnarray}
	P_f &=& \sum_{i=1}^{V_f} l_i\\
	A_f &=& \frac{1}{2} \sum_{i=1}^{V_f-1}\sum_{j>i}^{V_f-1} (l_j^x l_i^y - l_j^y l_i^x) \; ,
	\end{eqnarray}
where $\vec{l}_i = \vec{r}_{i+1}-\vec{r}_{i}+\vec{u}_{i+1}-\vec{u}_{i}$ is the length of edge $i$ of face $f$ the same way as defined in Sec.~\ref{sec:model}. Thus the expansion on perimeter and area can be arranged in orders of $u$ as:
	\begin{equation}
	l_i = l_i^{(0)}+ l_i^{(1)}+l_i^{(2)}+\order{l_i^{(3)}}
	\end{equation}
where 
	\begin{eqnarray}
	l_i^{(0)} &=& \vec{r}_{i+1}-\vec{r}_{i}\\ \label{edge0}
	l_i^{(1)} &=& \mathbf{u}\cdot \nabla l_i = \qty(\vec{u}_{i+1}-\vec{u}_i)\cdot \hat{n}_i\\ \label{edge1}
	l_i^{(2)} &=& \mathbf{u}\cdot \nabla \nabla^T l_i
	\cdot \mathbf{u}^T
	=\frac{1}{2l_i}\qty|\hat{n}_i \times \qty(\vec{u}_{i+1}-\vec{u}_i)|^2 \label{edge2}
	\end{eqnarray}
with $\hat{n}_i$ being the unit vector of $\vec{l}^{(0)}_i$, which is the bond direction before displacements. 
Subjecting Eq.~\eqref{edge0} - \eqref{edge2} into Eq.~\eqref{expansionVM} and \eqref{expansionAT} allows us to find explicit expressions in terms of the displacements $\vb{u}$.

For the VM, the 1st quadratic term in Eq.~\eqref{expansionVM} can be rearranged to become a sum over faces
    \begin{equation}
    \sum_{\expval{ij}}\frac{K_P}{2}\qty[\qty(\vb{u}\cdot \nabla \qty(P_{a}+P_{b}))\qty(\vb{u}\cdot \nabla R_{ij})] = \sum_{f}\frac{K_P}{2} \qty[\sum_{i}^{n} l^{(1)}_i]^2 .
    \end{equation}
The 2nd term after the edge to face summation rearrangements becomes
    \begin{equation}
    \sum_{\expval{ij}}\qty(T_a+T_b)\qty(\vb{u}\cdot \nabla \nabla^T R_{ij} \cdot \vb{u}^T) = \sum_f T_f \sum_{i=1}^n l^{(2)}_i
    \end{equation}
where $T_f$ is the cortical tension on cell $f$. 

It is obvious now that all $\order{u^2}$ terms in the elastic energy of the VM can be arranged into these complete square terms, the total number of which is equal to $F+E$. 
For a ZM which leaves the elastic energy zero, each of the complete square terms need to vanish. We thus arrive at ZM conditions for the VM
    \begin{eqnarray}
    \sum_{i}^{V_f} \qty(\vec{u}_{i+1}-\vec{u}_i)\cdot \hat{n}_i = 0\\ \label{rot}
    \hat{n}_i \times \qty(\vec{u}_{i+1}-\vec{u}_i) = 0 \label{peri}
    \end{eqnarray}
which are Eq.~\eqref{ZMPeri} and \eqref{ZMRot} in the main text. {These constraints are generically linearly independent unless the geometry is fine-tuned such that a singularity arises.}

We can do a similar analysis for the ATN, where we find that the 1st term in Eq.~\eqref{expansionAT}
    \begin{equation}
    \sum_{\expval{ij}}\frac{T_{ij}}{2}\qty(\vb{u}\cdot \nabla \nabla^T R_{ij} \cdot \vb{u}^T)
    \end{equation}
result in exactly the same ZM condition as Eq.~\eqref{rot}. The 2nd term
    \begin{equation}
    \sum_f \frac{K_A}{2}   \qty(\vb{u}\cdot \nabla A_f)^2
    \end{equation}
has the completed square on $\vb{u}\cdot \nabla A_f$, which leads to  the ZM condition 
described in Eq.~\eqref{ZMArea} of the main text. Interestingly, the 3rd term in Eq.~\ref{expansionAT}
    \begin{equation}
    \frac{1}{2}  \Pi_f \qty(\vb{u}\cdot \nabla \nabla^T A_{f}\cdot \vb{u}^T)
    \end{equation}
 only depends on the boundary displacements, because it becomes the variation of the total area of the whole sheet after summing over all faces.  Thus, this term vanishes automatically for any internal vertex and does not provide a new constraint.  
The total number of constraints in the ATN is also $F+E$, placing it at the Maxwell point.

\section{Compatibility Matrix}\label{APP:CM}
\subsection{Compatibility matrix of the ATN}
The compatibility matrix for the ATN can be constructed according to the ZM conditions in Eq.~\eqref{ZMRot} and Eq.~\eqref{ZMArea}. 
For the unit cell construction in Fig.~\ref{Unit_Cell}, the momentum space compatibility matrix $\vb{C}(\vb{k})$ is an $8\times 8$ matrix because we have 4 sites, 2 faces and 6 edges in a unit cell. From the condition Eq.~\eqref{ZMRot}, we have 6 constraints for  ZMs in the ATN,
	\begin{align}
	&\hat{n}_1 \times \qty(\vec{u}_2 - e^{ik_1}\vec{u}_1) = \vb{e}^\perp_1 \label{EQ:Ckbond1}\\ 
	&\hat{n}_2 \times \qty(\vec{u}_1 - \vec{u}_2) = \vb{e}^\perp_2 \label{EQ:Ckbond2}\\
	&\hat{n}_3 \times \qty(e^{-ik_2}\vec{u}_4 - \vec{u}_1) = \vb{e}^\perp_3 \label{EQ:Ckbond3}\\
	&\hat{n}_4 \times \qty(\vec{u}_3 - e^{-ik_2}\vec{u}_4) = \vb{e}^\perp_4 \label{EQ:Ckbond4}\\
	&\hat{n}_5 \times \qty(\vec{u}_4 - \vec{u}_3) = \vb{e}^\perp_5 \label{EQ:Ckbond5}\\
	&\hat{n}_6 \times \qty(e^{i(k_1-k_2)}\vec{u}_2 - \vec{u}_3) = \vb{e}^\perp_6 \label{EQ:Ckbond6}
	\end{align}	
And from the area conditions in Eq.~\eqref{ZMArea}, we have 2 more constraints for ZMs in the ATN,
	\begin{equation} \label{EQ:Ckarea1}
	\begin{split}
	&\qty(\vec{u}_2-e^{ik_1}\vec{u}_1)\times \qty(\vec{l}_2+\vec{l}_3)+\qty(\vec{u}_1 - \vec{u}_2)\times \qty(-\vec{l}_1+\vec{l}_3)\\
	&-\qty(e^{-ik_2}\vec{u}_4 - \vec{u}_1)\times \qty(\vec{l}_2-\vec{l}_1) +\qty(\vec{u}_3 - e^{-ik_2}\vec{u}_4)\times \qty(\vec{l}_5-\vec{l}_3)\\
	 &+\qty(\vec{u}_4 - \vec{u}_3)\times \qty(-\vec{l}_3-\vec{l}_4)\\
	 &-\qty(e^{ik_1}\vec{u}_1 - \vec{u}_4)\times \qty(\vec{l}_4+\vec{l}_5) =\Delta A_1
	\end{split}
	\end{equation}
and
	\begin{equation} \label{EQ:Ckarea2}
	\begin{split}
	&\qty(\vec{u}_4-e^{ik_1}\vec{u}_3)\times \qty(\vec{l}_6-\vec{l}_5)+\qty(\vec{u}_3 - \vec{u}_4)\times \qty(\vec{l}_6+\vec{l}_4)\\
	&-\qty(e^{i(k_1-k_2)}\vec{u}_2 - \vec{u}_3)\times \qty(-\vec{l}_4-\vec{l}_5)\\
	 &+\qty(e^{ik_2}\vec{u}_1 - e^{i(k_1-k_2)}\vec{u}_2)\times \qty(-\vec{l}_2-\vec{l}_6)\\
	 &+\qty( e^{ik_2}\vec{u}_2 -  e^{ik_2}\vec{u}_1)\times \qty(\vec{l}_1-\vec{l}_6)\\
	 &-\qty(e^{ik_1}\vec{u}_3 -  e^{ik_2}\vec{u}_2)\times \qty(-\vec{l}_1-\vec{l}_2) =\Delta A_2
	\end{split}
	\end{equation}

Putting the 8 equations from Eq.~\eqref{EQ:Ckbond1} - Eq.~\eqref{EQ:Ckarea2} together results in the compatibility matrix of size $8 \times 8$ in the basis of $\qty{u^x_i, u^y_i}$ where $i=1,2,\cdots,4$ with each constraint in each row, and each degrees of freedom in each column. The determinant of this matrix is used to compute the topological polarization.

\subsection{Compatibility matrix of the VM}
Similarly, we can construct the compatibility matrix for the VM using Eq.~\eqref{ZMRot} and Eq.~\eqref{ZMPeri}. The bond rotation constraint Eq.~\eqref{ZMRot} have the same 6 equations as shown in Eqs.~(\ref{EQ:Ckbond1} - \ref{EQ:Ckbond6}). However, the perimeter conservation Eq.~\eqref{ZMPeri} gives us 2 constraints,
    \begin{equation} \label{EQ: Ckperi1}
    \begin{split}
    &\qty(\vec{u}_2-e^{ik_1}\vec{u}_1) \cdot \vec{l}_1 + \qty(\vec{u}_1 - \vec{u}_2)\cdot \vec{l}_2
	\qty(e^{-ik_2}\vec{u}_4 - \vec{u}_3)\cdot \vec{l}_3\\ 
	&+\qty(\vec{u}_3 - e^{-ik_2}\vec{u}_4)\cdot \vec{l}_4
	 +\qty(\vec{u}_4 - \vec{u}_3)\cdot \vec{l}_5+\qty(e^{ik_1}\vec{u}_1 - \vec{u}_4)\cdot \vec{l}_6 =\Delta P_1
	\end{split}
    \end{equation}
    \begin{equation} \label{EQ: Ckperi2}
    \begin{split}
    &\qty(\vec{u}_4-e^{ik_1}\vec{u}_3) \cdot \vec{l}_1 + \qty(\vec{u}_3 - \vec{u}_4)\cdot \vec{l}_2
	\qty(e^{i(k_1-k_2)}\vec{u}_2 - \vec{u}_3)\cdot \vec{l}_3\\ 
	&+\qty(e^{ik_2}\vec{u}_1 - e^{i(k_1-k_2)}\vec{u}_2)\cdot \vec{l}_4
	 +\qty(e^{ik_2}\vec{u}_2 -  e^{ik_2}\vec{u}_1)\cdot \vec{l}_5+\qty(e^{ik_1}\vec{u}_3 -  e^{ik_2}\vec{u}_2)\cdot \vec{l}_6 =\Delta P_2.
	 \end{split}
    \end{equation}
The $8 \times 8$ compatibility matrix of the VM is constructed using Eqs.~(\ref{EQ:Ckbond1} - \ref{EQ:Ckbond6}), \eqref{EQ: Ckperi1}, and \eqref{EQ: Ckperi2} again in the basis of $\qty{u^x_i, u^y_i}$.

\section{Topological Phase Diagram of Networks Close to Critical Configurations Along  $\vec{a}_1$} \label{APP:Pa1}
In Fig.~\ref{bm} we show two critical configurations with bonds forming straight lines,  allowing bulk ZMs where cells translate along strips along $\vec{a}_2$ (a) or $\vec{a}_1$ (b).  In the main text, we discussed the phase diagram around the critical configuration (a).  

We have done similar analysis of configurations around critical configuration (b), and the results are shown in Fig.~\ref{pd1}.  Similarly, the sheet is unpolarized when all cells are convex.  When the cells become concave, configurations of $\vec{R_T}=\pm \vec{a}_1$, as well as a region with Weyl points show up.  We show some examples of the geometry of these phases in Fig.~\ref{FIG:pd1examples}.

It is interesting to note here that because 4 constraints per unit cell are removed when a vertical cut is introduced on the lattice to generate an open boundary along $\vec{a}_2$ (so as to show the topological polarization along $\vec{a}_1$), 4 instead of 2 ZMs are generated per unit cell.  Therefore, topological polarization $\vec{R_T}=\pm \vec{a}_1$ indicates that the ratio of ZMs at the left and right boundary have ratios of $1:3$ or $3:1$ instead of $0:4$ or $4:0$.  As a result, we do not observe any boundary becoming completely ZM free in this case, unlike the phase diagram we discussed in the main text, where the top or the bottom boundaries can be free of ZMs.

\begin{figure}[h!]
	\centering
    \includegraphics[scale=0.45]{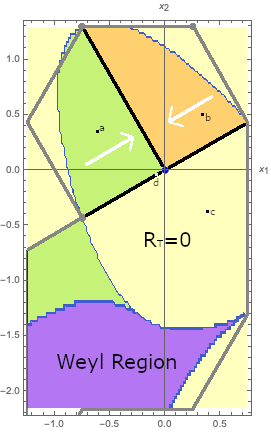}
    \caption{Phase diagram for changing site 2 coordinate in Case 1 with the same representation style as Fig.~\ref{pd2}. The gray boundary labels the outbound of the unit cell with the 3 stationary sites besides vertex 2. The thick black line marks  critical  configurations, and 5 different topological phases are observed.  The yellow region is un-polarized. The cyan, and red regions are topologically polarized with $\vec{R}_T$ along $\vec{a}_1$, and $-\vec{a}_1$ respectively, where the white arrows mark $\vec{R}_T$.  In the purple region the lattice displays Weyl points and thus topologically protected bulk floppy modes.  Four representative configurations of these regions (marked by black dots) are shown in Fig.~\ref{FIG:pd1examples}.}
    \label{pd1}
\end{figure}
\begin{figure}[h!]
	\centering
    \includegraphics[scale=0.28]{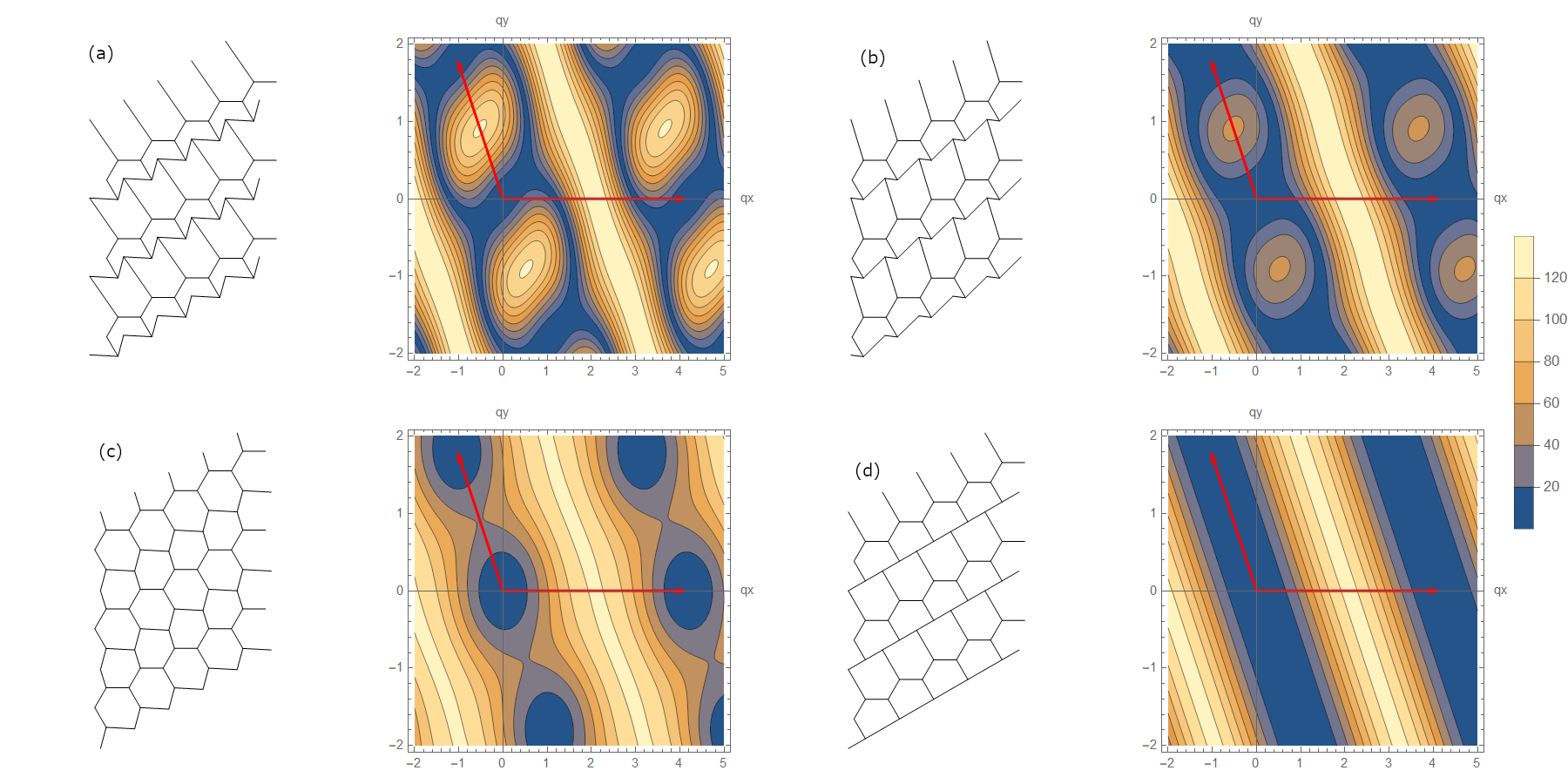}
    \caption{Representative examples of cell sheet lattices in different regions of the phase diagram (Fig.~\ref{pd1}) with the same style as Fig.~\ref{FIG: Example_Network}. (a): A polarized lattice with $\vec{R}_T = \vec{a}_1$. (b): A polarized lattice with $\vec{R}_T= -\vec{a}_1$. (c): An unpolarized lattice. (d): A lattice at critical configuration.}
    \label{FIG:pd1examples}
\end{figure}

\section{Transfer Matrix for Disordered Cell Sheets} \label{APP:TM}
In this appendix we develop a transfer matrix method for ZMs in cell sheets, which can  be applied to \emph{disordered} cell sheets to conveniently derive the ZM at given boundary conditions.  The transfer matrix for the VM and the ATN can be derived in similar ways, thus we show both derivations in this appendix.

In this transfer matrix construction, we assume that each cell is a hexagon (of arbitrary shape) and each vertex has three edges meeting at it, so the sheet still has the topology of a honeycomb lattice, but no periodicity is required for the shapes of the cells.   The constructed transfer matrix will enable us to derive the ZM displacements of the three ``outgoing'' edges from the ZM displacements of the three ``incoming'' edges.  Therefore by propagating this transfer matrix through the whole cell sheet, where each hexagonal cell has three in-flux and three out-flux, we can compute the ZM of the whole sheet.

\begin{figure}[h!]
\includegraphics[width=0.3\textwidth]{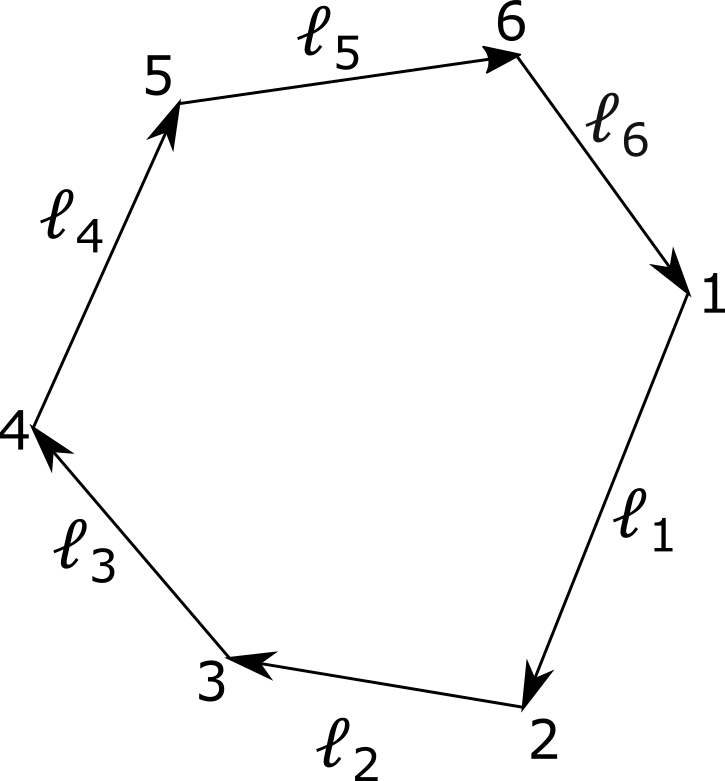}
\caption{The convention used in establishing the Transfer Matrix, sites and edges are labeled as in the figure, and edge directions are chosen to be in the clockwise direction.}
\label{hexcell}
\end{figure}

To derive this transfer matrix method,  we consider one cell and establish the edge conventions as shown in Fig.~\ref{hexcell}.  Same as in Eq.~\eqref{ZMArea}, we define $\vec{\mathcal{U}}_i = \vec{u}_{i+1}-\vec{u}_i$ as the relative displacement between the neighboring vertices.  Because the ZM cannot rotate the edges [Eq.\eqref{ZMRot}], these vectors $\vec{U}_i$ only have components parallel to the original edge direction
	\begin{equation} \label{EQ:TM1}
	\mathcal{U}^\parallel_i = \vec{\mathcal{U}} \cdot \hat{l}_{i,i+1},
	\end{equation}
where $\hat{l}_{i,i+1}$ is the direction of the edge connecting vertices $i$ and $i+1$.  

At each cell, we assume there are three known displacements $\{\mathcal{U}^\parallel_1,\mathcal{U}^\parallel_2,\mathcal{U}^\parallel_3\}$, and we will derive three unknown displacements $\{\mathcal{U}^\parallel_4,\mathcal{U}^\parallel_5,\mathcal{U}^\parallel_6\}$.  From the fact that the hexagonal cell has to remain closed, we have
	\begin{equation} \label{EQ:TM2}
	\sum_{i=1}^{V_f}  \mathcal{U}^\parallel_i \hat{l}_{i,i+1} =0.
	\end{equation}
This gives us two equations, because it is a vectorial equation.  
One more equation for ZMs comes from the perimeter conservation condition in Eq.~\eqref{ZMPeri} in the VM
    \begin{equation} \label{EQ:TM3}
    \sum_{i=1}^{V_f} \vec{\mathcal{U}} \cdot \hat{l}_{i,i+1} =  \sum_{i=1}^{V_f} \mathcal{U}^\parallel_i = 0
    \end{equation}
and from the area preservation condition in Eq.~\eqref{ZMArea} in the ATN.
	\begin{equation} \label{EQ:TM4}
	\sum_{i=1}^{V_f-1}\sum_{j>i}^{V_f-1} \qty(\vec{\mathcal{U}}_j\cross \vec{\mathcal{L}}_{i} - \vec{\mathcal{U}}_i\cross \vec{\mathcal{L}}_{j}) =0 . 
	\end{equation}

Equations~\eqref{EQ:TM2} - Eq.~\eqref{EQ:TM4} allow us to write a transfer matrix $\vb{M}$ for both models such that
	\begin{equation}
	\vb{M} \cdot \mqty(\mathcal{U}_1^\parallel\\\mathcal{U}_2^\parallel\\\mathcal{U}_3^\parallel) =  \mqty(\mathcal{U}_4^\parallel\\\mathcal{U}_5^\parallel\\\mathcal{U}_6^\parallel)
	\end{equation}

The transfer matrix $\vb{M}$ takes the form of a square non-symmetric matrix for both models. In the VM,
    \begin{equation} \label{EQ:TMVM}
    \vb{M}_{VM}=-\mqty(1&1&1&\\\cos{\theta_4}&\cos{\theta_5}&\cos{\theta_6}\\\sin{\theta_4}&\sin{\theta_5}&\sin{\theta_6})^{-1} \cdot
    \mqty(1&1&1&\\\cos{\theta_1}&\cos{\theta_2}&\cos{\theta_3}\\\sin{\theta_1}&\sin{\theta_2}&\sin{\theta_3})
    \end{equation}
where $\theta_i$ are the angles of the edges $\mathcal{L}_{i,i+1}$ in the Cartesian coordinate system. The transfer matrix $\vb{M}$ for the ATN has a similar form as Eq.~\eqref{EQ:TMVM}, with the elements in the first row replaced by the terms given by Eq.~\eqref{EQ:TM4}.
	\begin{equation} \label{EQ:TMATN}
	\vb{M}_{ATN} = - \smqty(\cos \theta_4 \qty(\mathcal{L}_5^{y}+\mathcal{L}_6^{y})- \sin \theta_4 \qty(\mathcal{L}_5^{x}+\mathcal{L}_6^{x}) & \cos \theta_5 \qty(\mathcal{L}_6^{y}-\mathcal{L}_4^{y})- \sin \theta_5 \qty(\mathcal{L}_6^{x}-\mathcal{L}_4^{x}) & -\cos \theta_6 \qty(\mathcal{L}_4^{y}+\mathcal{L}_5^{y})+ \sin \theta_6 \qty(\mathcal{L}_4^{x}+\mathcal{L}_5^{x})\\
	 \cos\theta_4 & \cos\theta_5 & \cos \theta_6\\
	\sin\theta_4 & \sin\theta_5 & \sin \theta_6
	)^{-1}
	\end{equation}
	\begin{equation*}
	\cdot\smqty(\cos \theta_1 \qty(\mathcal{L}_2^{y}+\mathcal{L}_3^{y})- \sin \theta_1 \qty(\mathcal{L}_2^{x}+\mathcal{L}_3^{x}) & \cos \theta_2 \qty(\mathcal{L}_3^{y}-\mathcal{L}_1^{y})- \sin \theta_2 \qty(\mathcal{L}_3^{x}-\mathcal{L}_1^{x}) & -\cos \theta_3 \qty(\mathcal{L}_1^{y}+\mathcal{L}_2^{y})+ \sin \theta_3 \qty(\mathcal{L}_1^{x}+\mathcal{L}_2^{x})\\
	\cos\theta_1 & \cos\theta_2 & \cos \theta_3\\
	\sin\theta_1 & \sin\theta_2 & \sin \theta_3
	)
	\end{equation*}
These transfer matrices can be used to propagate the ZM across the whole sheet cell by cell from given boundary conditions, as shown in Fig.~\ref{FIG:TM_Propagation}.

\begin{figure}[h!]
    \centering
    \includegraphics[scale=0.2]{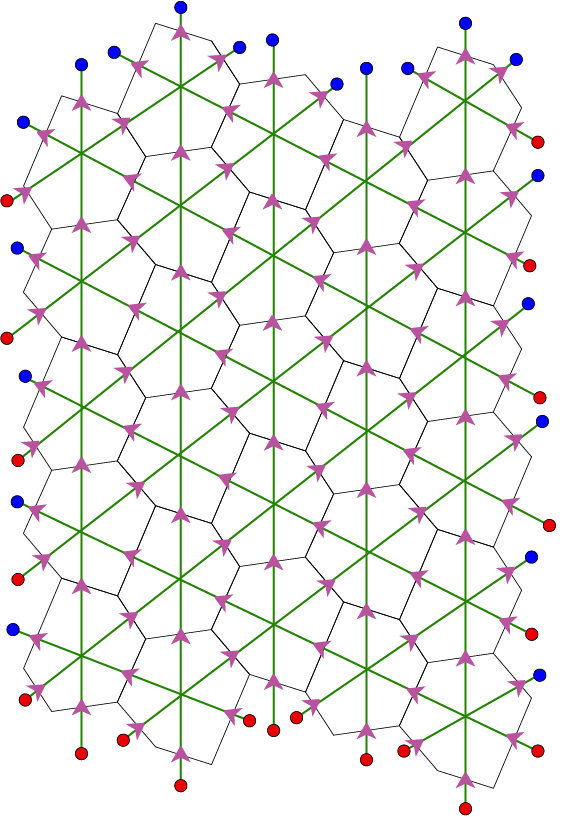}
    \caption{An illustration of how the transfer matrix can use the "incoming" edges to solve for the "outgoing" edges on a sheet of hexagonal cells for ZMs. The red dots label the ``incoming" edges, where $\mathcal{U}_i^\parallel$ are given by boundary conditions, 
    and the blue dots represent the ``outgoing" edges where $\mathcal{U}_i^\parallel$  are calculated. 
    Across each cell, the transfer matrix allows us to find out the ZM at the three outgoing edges as functions of the ZM at the three incoming edges, and the direction of the ZM solution propagation is labeled by the magenta arrows across cell edges. The  choice of the in and out directions is not unique on the sheet, and depends on which boundaries are fixed.  The total number of incoming edges (red dots) is equal to the total number of ZMs of the whole sheet, so determining $\mathcal{U}_i^\parallel$ at these edges determines the ZM of the whole sheet.
     }
    \label{FIG:TM_Propagation}
\end{figure}

To study topological mechanics, we again start from the bulk ZM configuration shown in Fig.~\ref{bm}(a), where
$\vec{\mathcal{U}}_1\parallel\vec{\mathcal{U}}_3\parallel\vec{\mathcal{U}}_4\parallel\vec{\mathcal{U}}_6$.  In this configuration, both $\vb{M}_{VM}$ and $\vb{M}_{ATN}$ yield one ZM with eigenvalue $\lambda_1=-1$ with a corresponding eigenvector $\vec{\nu}_1=\mqty(1\\0\\1)$ for the $\mathcal{U}^\parallel_i$.  This is clearly the bulk ZM depicted in Fig.~\ref{bm}(a), where edges 2 and 5 shift vertically.

However, the other two eigenmodes of this $3\times3$ transfer matrix do not correspond to simple decompositions of other edges of this cell.  
Specifically in ATN, one of these two modes represents vertical shifts of neighboring columns with an eigenvalue $\lambda_2 = 1$, whereas the other one represents a horizontal broadening or narrowing of the network that has a corresponding eigenvalue $\lambda_3 = \frac{\qty|\mathcal{L}_5|}{\qty|\mathcal{L}_2|}$, and this mode captures the "breathing" mode discussed in the main text.
This differs from other simpler cases of transfer matrices for topological mechanics where modes symmetrically separate, making it transparent to study ZM decay in different directions in those systems~\cite{Zhou2018,Zhou2019}.

For the rest of this appendix, 
we introduce perturbations to the vertex positions and examine how the eigenvalues of the transfer matrix change.  In particular, we focus on the first mode which has eigenvalue $\lambda_1=-1$ at the critical configuration.  The sign of its first order correction $\delta \lambda_1$ indicates the  directions of decay in the ZM.

We use  first order perturbation theory to find $\delta \lambda $ as a function of the geometric perturbation of the vertex positions. However, due to the non-symmetric nature of the transfer matrix, the first order perturbation method needs a slight modification from the usual perturbation theory because the left and right eigenvectors of the matrix are not identical. 

In  first order perturbation theory for symmetric matrices, the perturbation to the eigenvalues take the form $E^{(1)} = \expval{\delta E}{\psi^{(0)}}$.  However for non-symmetric matrices, it takes the form $$\delta \lambda_i= \frac{\vec{\mu^\intercal}_i^0 \vb{\delta M} \vec{\nu}_i^0}{\vec{\mu^\intercal}_i^0 \vec{\nu}_i^0}$$ where $\vec{\mu^\intercal}_i^0, \vec{\nu}_i^0$ are the left and right eigenvectors of the unperturbed matrix $\vb{M}$. The derivation of this form is supplied as the following.

With some small  geometric change from the critical configuration, the  transfer matrix $\vb{M}$ can be written as
	\begin{equation}
	\vb{M} = \vb{M}^0+\vb{\delta M} ,
	\end{equation}
the  eigenvalues $\lambda_i$ can be expanded to the first order as 
	\begin{equation}
	\lambda_i \approx \lambda^0_{i} +\delta \lambda
	\end{equation}
and the right eigenvectors $\vec{\nu}_i$ can be expanded to the first order as
	\begin{equation}
	\vec{\nu}_i \approx \vec{\nu}^0_{i} +\vec{\delta \nu}
	\end{equation}
where $\vb{M}^0$, $\lambda^0_i$, and $\vec{\nu}^0_i$ are the transfer matrix and its eigenvalues and eigenvectors when the geometry is at the critical configuration, so that we know $\vb{M}^0 \vec{\nu}^0_i = \lambda^0_i \vec{\nu}^0_i$.

With these expansions, we have  
	\begin{equation}
	\qty(\vb{M}^0+\vb{\delta M}) \qty(\vec{\nu}^0_i+\vec{\delta \nu}_i) = \qty(\lambda^0_i + \delta \lambda_i) \qty(\vec{\nu}^0_i+\vec{\delta \nu}_i).
	\end{equation}
Multiplying out the terms and keep  to the first order, we get
	\begin{equation}
	\vb{M}^0 \vec{\nu}^0_i + \vb{M}^0 \vec{\delta \nu}_i + \vb{\delta M} \vec{\nu}_i^0 =  \lambda_i^0 \vec{\nu}_i^0+\lambda_i \vec{\delta \nu}_i + \delta \lambda_i \vec{\nu}_i^0 .
	\end{equation}
The first terms cancel on both sides, so it becomes
	\begin{equation}
	\vb{M}^0 \vec{\delta \nu}_i + \vb{\delta M} \vec{\nu}_i^0 = \lambda_i \vec{\delta \nu}_i + \delta \lambda_i \vec{\nu}_i^0 . \label{Expansion}
	\end{equation}
Now suppose  $\vec{\mu}^0_i$ is the left eigenvector of $\vb{M}^0$ with the eigenvalue $\lambda_i$, such that $\vec{\mu^\intercal}_i^0 \vb{M}^0 = \vec{\mu^\intercal}_i^0 \lambda_i$, or equivalently, $\vb{M^\intercal} \vec{\mu}_i^0 = \lambda_i^0 \vec{\mu}_i^0$.   
Dotting this left eigenvector on both sides of the equation gives us
	\begin{equation}
	\vec{\mu^\intercal}_i^0 \vb{M}^0 \vec{\delta \nu}_i + \vec{\mu^\intercal}_i^0 \vb{\delta M} \vec{\nu}_i^0 =  \vec{\mu^\intercal}_i^0 \lambda_i \vec{\delta \nu}_i + \vec{\mu^\intercal}_i^0 \delta \lambda_i \vec{\nu}_i^0
	\end{equation}
Now we can cancel out the first term on both sides again based on the property of the left eigenvectors $\vec{\mu}_i^0$, so we are left with 
	\begin{equation}
	\vec{\mu^\intercal}_i^0 \vb{\delta M} \vec{\nu}_i^0 = \vec{\mu^\intercal}_i^0 \delta \lambda_i \vec{\nu}_i^0
	\end{equation}
Rearranging the equation, we have
	\begin{equation}
	\delta \lambda_i = \frac{\vec{\mu^\intercal}_i^0 \vb{\delta M} \vec{\nu}_i^0}{\vec{\mu^\intercal}_i^0 \vec{\nu}_i^0}
	\end{equation}
which is the first order correction to the eigenvalue $\lambda_i$ of the transfer matrix.

This perturbation theory can potentially be used to study how ZMs exponentially grow or decay in disordered cell sheets.  We applied this method to the periodic lattices we studied in the main text, and  the results are consistent between the momentum space calculation described in the main text and the transfer matrix calculation.  As we choose lattices in each topological phase in Fig.~\ref{pd2}, the sign changes of $\delta \lambda$ agrees with the winding number jumps.
\end{widetext}

\bibliography{VM_Draft}

\begin{thebibliography}{48}%
\makeatletter
\providecommand \@ifxundefined [1]{%
 \@ifx{#1\undefined}
}%
\providecommand \@ifnum [1]{%
 \ifnum #1\expandafter \@firstoftwo
 \else \expandafter \@secondoftwo
 \fi
}%
\providecommand \@ifx [1]{%
 \ifx #1\expandafter \@firstoftwo
 \else \expandafter \@secondoftwo
 \fi
}%
\providecommand \natexlab [1]{#1}%
\providecommand \enquote  [1]{``#1''}%
\providecommand \bibnamefont  [1]{#1}%
\providecommand \bibfnamefont [1]{#1}%
\providecommand \citenamefont [1]{#1}%
\providecommand \href@noop [0]{\@secondoftwo}%
\providecommand \href [0]{\begingroup \@sanitize@url \@href}%
\providecommand \@href[1]{\@@startlink{#1}\@@href}%
\providecommand \@@href[1]{\endgroup#1\@@endlink}%
\providecommand \@sanitize@url [0]{\catcode `\\12\catcode `\$12\catcode
  `\&12\catcode `\#12\catcode `\^12\catcode `\_12\catcode `\%12\relax}%
\providecommand \@@startlink[1]{}%
\providecommand \@@endlink[0]{}%
\providecommand \url  [0]{\begingroup\@sanitize@url \@url }%
\providecommand \@url [1]{\endgroup\@href {#1}{\urlprefix }}%
\providecommand \urlprefix  [0]{URL }%
\providecommand \Eprint [0]{\href }%
\providecommand \doibase [0]{https://doi.org/}%
\providecommand \selectlanguage [0]{\@gobble}%
\providecommand \bibinfo  [0]{\@secondoftwo}%
\providecommand \bibfield  [0]{\@secondoftwo}%
\providecommand \translation [1]{[#1]}%
\providecommand \BibitemOpen [0]{}%
\providecommand \bibitemStop [0]{}%
\providecommand \bibitemNoStop [0]{.\EOS\space}%
\providecommand \EOS [0]{\spacefactor3000\relax}%
\providecommand \BibitemShut  [1]{\csname bibitem#1\endcsname}%
\let\auto@bib@innerbib\@empty
\bibitem [{\citenamefont {Schöck}\ and\ \citenamefont
  {Perrimon}(2002)}]{Schock2002}%
  \BibitemOpen
  \bibfield  {author} {\bibinfo {author} {\bibfnamefont {F.}~\bibnamefont
  {Schöck}}\ and\ \bibinfo {author} {\bibfnamefont {N.}~\bibnamefont
  {Perrimon}},\ }\bibfield  {title} {\bibinfo {title} {Molecular mechanisms of
  epithelial morphogenesis},\ }\href
  {https://doi.org/10.1146/annurev.cellbio.18.022602.131838} {\bibfield
  {journal} {\bibinfo  {journal} {Annual Review of Cell and Developmental
  Biology}\ }\textbf {\bibinfo {volume} {18}},\ \bibinfo {pages} {463}
  (\bibinfo {year} {2002})}\BibitemShut {NoStop}%
\bibitem [{\citenamefont {Solnica-Krezel}\ and\ \citenamefont
  {Sepich}(2012)}]{Sepich2012}%
  \BibitemOpen
  \bibfield  {author} {\bibinfo {author} {\bibfnamefont {L.}~\bibnamefont
  {Solnica-Krezel}}\ and\ \bibinfo {author} {\bibfnamefont {D.~S.}\
  \bibnamefont {Sepich}},\ }\bibfield  {title} {\bibinfo {title} {Gastrulation:
  Making and shaping germ layers},\ }\href
  {https://doi.org/10.1146/annurev-cellbio-092910-154043} {\bibfield  {journal}
  {\bibinfo  {journal} {Annual Review of Cell and Developmental Biology}\
  }\textbf {\bibinfo {volume} {28}},\ \bibinfo {pages} {687} (\bibinfo {year}
  {2012})}\BibitemShut {NoStop}%
\bibitem [{\citenamefont {Fristrom}(1988)}]{Fristrom1988}%
  \BibitemOpen
  \bibfield  {author} {\bibinfo {author} {\bibfnamefont {D.}~\bibnamefont
  {Fristrom}},\ }\bibfield  {title} {\bibinfo {title} {The cellular basis of
  epithelial morphogenesis. a review},\ }\href
  {https://doi.org/10.1016/0040-8166(88)90015-8} {\bibfield  {journal}
  {\bibinfo  {journal} {Tissue and Cell}\ }\textbf {\bibinfo {volume} {20}},\
  \bibinfo {pages} {645} (\bibinfo {year} {1988})}\BibitemShut {NoStop}%
\bibitem [{\citenamefont {Leptin}(2005)}]{Leptin2005}%
  \BibitemOpen
  \bibfield  {author} {\bibinfo {author} {\bibfnamefont {M.}~\bibnamefont
  {Leptin}},\ }\bibfield  {title} {\bibinfo {title} {Gastrulation movements:
  The logic and the nuts and bolts},\ }\href
  {https://doi.org/10.1016/j.devcel.2005.02.007} {\bibfield  {journal}
  {\bibinfo  {journal} {Developmental Cell}\ }\textbf {\bibinfo {volume} {8}},\
  \bibinfo {pages} {305} (\bibinfo {year} {2005})}\BibitemShut {NoStop}%
\bibitem [{\citenamefont {Colas}\ and\ \citenamefont
  {Schoenwolf}(2001)}]{Colas2001}%
  \BibitemOpen
  \bibfield  {author} {\bibinfo {author} {\bibfnamefont {J.~F.}\ \bibnamefont
  {Colas}}\ and\ \bibinfo {author} {\bibfnamefont {G.~C.}\ \bibnamefont
  {Schoenwolf}},\ }\bibfield  {title} {\bibinfo {title} {Towards a cellular and
  molecular understanding of neurulation},\ }\href
  {https://doi.org/10.1002/dvdy.1144} {\bibfield  {journal} {\bibinfo
  {journal} {Developmental Dynamics}\ }\textbf {\bibinfo {volume} {221}},\
  \bibinfo {pages} {117} (\bibinfo {year} {2001})}\BibitemShut {NoStop}%
\bibitem [{\citenamefont {Trichas}\ \emph {et~al.}(2012)\citenamefont
  {Trichas}, \citenamefont {Smith}, \citenamefont {White}, \citenamefont
  {Wilkins}, \citenamefont {Watanabe}, \citenamefont {Moore}, \citenamefont
  {Joyce}, \citenamefont {Sugnaseelan}, \citenamefont {Rodriguez},
  \citenamefont {Kay}, \citenamefont {Baker}, \citenamefont {Maini},\ and\
  \citenamefont {Srinivas}}]{Trichas2012}%
  \BibitemOpen
  \bibfield  {author} {\bibinfo {author} {\bibfnamefont {G.}~\bibnamefont
  {Trichas}}, \bibinfo {author} {\bibfnamefont {A.~M.}\ \bibnamefont {Smith}},
  \bibinfo {author} {\bibfnamefont {N.}~\bibnamefont {White}}, \bibinfo
  {author} {\bibfnamefont {V.}~\bibnamefont {Wilkins}}, \bibinfo {author}
  {\bibfnamefont {T.}~\bibnamefont {Watanabe}}, \bibinfo {author}
  {\bibfnamefont {A.}~\bibnamefont {Moore}}, \bibinfo {author} {\bibfnamefont
  {B.}~\bibnamefont {Joyce}}, \bibinfo {author} {\bibfnamefont
  {J.}~\bibnamefont {Sugnaseelan}}, \bibinfo {author} {\bibfnamefont {T.~A.}\
  \bibnamefont {Rodriguez}}, \bibinfo {author} {\bibfnamefont {D.}~\bibnamefont
  {Kay}}, \bibinfo {author} {\bibfnamefont {R.~E.}\ \bibnamefont {Baker}},
  \bibinfo {author} {\bibfnamefont {P.~K.}\ \bibnamefont {Maini}},\ and\
  \bibinfo {author} {\bibfnamefont {S.}~\bibnamefont {Srinivas}},\ }\bibfield
  {title} {\bibinfo {title} {Multi-cellular rosettes in the mouse visceral
  endoderm facilitate the ordered migration of anterior visceral endoderm
  cells},\ }\href {https://doi.org/10.1371/journal.pbio.1001256} {\bibfield
  {journal} {\bibinfo  {journal} {PLoS Biology}\ }\textbf {\bibinfo {volume}
  {10}},\ \bibinfo {pages} {e1001256} (\bibinfo {year} {2012})}\BibitemShut
  {NoStop}%
\bibitem [{\citenamefont {Cowin}\ and\ \citenamefont {Doty}(2007)}]{Cowin2007}%
  \BibitemOpen
  \bibfield  {author} {\bibinfo {author} {\bibfnamefont {S.~C.}\ \bibnamefont
  {Cowin}}\ and\ \bibinfo {author} {\bibfnamefont {S.~B.}\ \bibnamefont
  {Doty}},\ }\href {https://doi.org/10.1007/978-0-387-49985-7} {\emph {\bibinfo
  {title} {Tissue Mechanics}}}\ (\bibinfo {year} {2007})\BibitemShut {NoStop}%
\bibitem [{\citenamefont {Guillot}\ and\ \citenamefont
  {Lecuit}(2013)}]{Guillot2013}%
  \BibitemOpen
  \bibfield  {author} {\bibinfo {author} {\bibfnamefont {C.}~\bibnamefont
  {Guillot}}\ and\ \bibinfo {author} {\bibfnamefont {T.}~\bibnamefont
  {Lecuit}},\ }\href {https://doi.org/10.1126/science.1235249} {\bibinfo
  {title} {{Mechanics of epithelial tissue homeostasis and morphogenesis}}}
  (\bibinfo {year} {2013})\BibitemShut {NoStop}%
\bibitem [{\citenamefont {Heller}\ and\ \citenamefont
  {Fuchs}(2015)}]{Heller2015}%
  \BibitemOpen
  \bibfield  {author} {\bibinfo {author} {\bibfnamefont {E.}~\bibnamefont
  {Heller}}\ and\ \bibinfo {author} {\bibfnamefont {E.}~\bibnamefont {Fuchs}},\
  }\href {https://doi.org/10.1083/jcb.201506106} {\bibinfo {title} {{Tissue
  patterning and cellular mechanics}}} (\bibinfo {year} {2015})\BibitemShut
  {NoStop}%
\bibitem [{\citenamefont {Lange}\ and\ \citenamefont
  {Fabry}(2013)}]{Lange2013}%
  \BibitemOpen
  \bibfield  {author} {\bibinfo {author} {\bibfnamefont {J.~R.}\ \bibnamefont
  {Lange}}\ and\ \bibinfo {author} {\bibfnamefont {B.}~\bibnamefont {Fabry}},\
  }\href {https://doi.org/10.1016/j.yexcr.2013.04.023} {\bibinfo {title} {{Cell
  and tissue mechanics in cell migration}}} (\bibinfo {year}
  {2013})\BibitemShut {NoStop}%
\bibitem [{\citenamefont {Landsberg}\ \emph {et~al.}(2009)\citenamefont
  {Landsberg}, \citenamefont {Farhadifar}, \citenamefont {Ranft}, \citenamefont
  {Umetsu}, \citenamefont {Widmann}, \citenamefont {Bittig}, \citenamefont
  {Said}, \citenamefont {Jülicher},\ and\ \citenamefont
  {Dahmann}}]{Landsberg2009}%
  \BibitemOpen
  \bibfield  {author} {\bibinfo {author} {\bibfnamefont {K.~P.}\ \bibnamefont
  {Landsberg}}, \bibinfo {author} {\bibfnamefont {R.}~\bibnamefont
  {Farhadifar}}, \bibinfo {author} {\bibfnamefont {J.}~\bibnamefont {Ranft}},
  \bibinfo {author} {\bibfnamefont {D.}~\bibnamefont {Umetsu}}, \bibinfo
  {author} {\bibfnamefont {T.~J.}\ \bibnamefont {Widmann}}, \bibinfo {author}
  {\bibfnamefont {T.}~\bibnamefont {Bittig}}, \bibinfo {author} {\bibfnamefont
  {A.}~\bibnamefont {Said}}, \bibinfo {author} {\bibfnamefont {F.}~\bibnamefont
  {Jülicher}},\ and\ \bibinfo {author} {\bibfnamefont {C.}~\bibnamefont
  {Dahmann}},\ }\bibfield  {title} {\bibinfo {title} {Increased cell bond
  tension governs cell sorting at the drosophila anteroposterior compartment
  boundary},\ }\href {https://doi.org/10.1016/j.cub.2009.10.021} {\bibfield
  {journal} {\bibinfo  {journal} {Current Biology}\ }\textbf {\bibinfo {volume}
  {19}},\ \bibinfo {pages} {1950} (\bibinfo {year} {2009})}\BibitemShut
  {NoStop}%
\bibitem [{\citenamefont {Aliee}\ \emph {et~al.}(2012)\citenamefont {Aliee},
  \citenamefont {Röper}, \citenamefont {Landsberg}, \citenamefont {Pentzold},
  \citenamefont {Widmann}, \citenamefont {Jülicher},\ and\ \citenamefont
  {Dahmann}}]{Aliee2012}%
  \BibitemOpen
  \bibfield  {author} {\bibinfo {author} {\bibfnamefont {M.}~\bibnamefont
  {Aliee}}, \bibinfo {author} {\bibfnamefont {J.~C.}\ \bibnamefont {Röper}},
  \bibinfo {author} {\bibfnamefont {K.~P.}\ \bibnamefont {Landsberg}}, \bibinfo
  {author} {\bibfnamefont {C.}~\bibnamefont {Pentzold}}, \bibinfo {author}
  {\bibfnamefont {T.~J.}\ \bibnamefont {Widmann}}, \bibinfo {author}
  {\bibfnamefont {F.}~\bibnamefont {Jülicher}},\ and\ \bibinfo {author}
  {\bibfnamefont {C.}~\bibnamefont {Dahmann}},\ }\bibfield  {title} {\bibinfo
  {title} {Physical mechanisms shaping the drosophila dorsoventral compartment
  boundary},\ }\href {https://doi.org/10.1016/j.cub.2012.03.070} {\bibfield
  {journal} {\bibinfo  {journal} {Current Biology}\ }\textbf {\bibinfo {volume}
  {22}},\ \bibinfo {pages} {967} (\bibinfo {year} {2012})}\BibitemShut
  {NoStop}%
\bibitem [{\citenamefont {Umetsu}\ \emph {et~al.}(2014)\citenamefont {Umetsu},
  \citenamefont {Aigouy}, \citenamefont {Aliee}, \citenamefont {Sui},
  \citenamefont {Eaton}, \citenamefont {Jülicher},\ and\ \citenamefont
  {Dahmann}}]{Umetsu2014}%
  \BibitemOpen
  \bibfield  {author} {\bibinfo {author} {\bibfnamefont {D.}~\bibnamefont
  {Umetsu}}, \bibinfo {author} {\bibfnamefont {B.}~\bibnamefont {Aigouy}},
  \bibinfo {author} {\bibfnamefont {M.}~\bibnamefont {Aliee}}, \bibinfo
  {author} {\bibfnamefont {L.}~\bibnamefont {Sui}}, \bibinfo {author}
  {\bibfnamefont {S.}~\bibnamefont {Eaton}}, \bibinfo {author} {\bibfnamefont
  {F.}~\bibnamefont {Jülicher}},\ and\ \bibinfo {author} {\bibfnamefont
  {C.}~\bibnamefont {Dahmann}},\ }\bibfield  {title} {\bibinfo {title} {Local
  increases in mechanical tension shape compartment boundaries by biasing cell
  intercalations},\ }\href {https://doi.org/10.1016/j.cub.2014.06.052}
  {\bibfield  {journal} {\bibinfo  {journal} {Current Biology}\ }\textbf
  {\bibinfo {volume} {24}},\ \bibinfo {pages} {1798} (\bibinfo {year}
  {2014})}\BibitemShut {NoStop}%
\bibitem [{\citenamefont {Kane}\ and\ \citenamefont
  {Lubensky}(2013)}]{Kane2013}%
  \BibitemOpen
  \bibfield  {author} {\bibinfo {author} {\bibfnamefont {C.~L.}\ \bibnamefont
  {Kane}}\ and\ \bibinfo {author} {\bibfnamefont {T.~C.}\ \bibnamefont
  {Lubensky}},\ }\bibfield  {title} {\bibinfo {title} {{Topological boundary
  modes in isostatic lattices}},\ }\href {https://doi.org/10.1038/nphys2835}
  {\bibfield  {journal} {\bibinfo  {journal} {Nature Physics}\ }\textbf
  {\bibinfo {volume} {10}},\ \bibinfo {pages} {39} (\bibinfo {year} {2013})},\
  \Eprint {https://arxiv.org/abs/1308.0554} {arXiv:1308.0554} \BibitemShut
  {NoStop}%
\bibitem [{\citenamefont {Lubensky}\ \emph {et~al.}(2015)\citenamefont
  {Lubensky}, \citenamefont {Kane}, \citenamefont {Mao}, \citenamefont
  {Souslov},\ and\ \citenamefont {Sun}}]{Lubensky2015}%
  \BibitemOpen
  \bibfield  {author} {\bibinfo {author} {\bibfnamefont {T.~C.}\ \bibnamefont
  {Lubensky}}, \bibinfo {author} {\bibfnamefont {C.~L.}\ \bibnamefont {Kane}},
  \bibinfo {author} {\bibfnamefont {X.}~\bibnamefont {Mao}}, \bibinfo {author}
  {\bibfnamefont {A.}~\bibnamefont {Souslov}},\ and\ \bibinfo {author}
  {\bibfnamefont {K.}~\bibnamefont {Sun}},\ }\bibfield  {title} {\bibinfo
  {title} {{Phonons and elasticity in critically coordinated lattices}},\
  }\href {https://doi.org/10.1088/0034-4885/78/7/073901} {\bibfield  {journal}
  {\bibinfo  {journal} {Reports on Progress in Physics}\ }\textbf {\bibinfo
  {volume} {78}},\ \bibinfo {pages} {73901} (\bibinfo {year} {2015})},\ \Eprint
  {https://arxiv.org/abs/1503.01324} {arXiv:1503.01324} \BibitemShut {NoStop}%
\bibitem [{\citenamefont {Mao}\ and\ \citenamefont {Lubensky}(2017)}]{Mao2017}%
  \BibitemOpen
  \bibfield  {author} {\bibinfo {author} {\bibfnamefont {X.}~\bibnamefont
  {Mao}}\ and\ \bibinfo {author} {\bibfnamefont {T.~C.}\ \bibnamefont
  {Lubensky}},\ }\bibfield  {title} {\bibinfo {title} {{Maxwell Lattices and
  Topological Mechanics}},\ }\bibfield  {journal} {\bibinfo  {journal} {Annual
  Review of Condensed Matter Physics}\ }\textbf {\bibinfo {volume} {9}},\ \href
  {https://doi.org/10.1146/annurev-conmatphys-033117-054235}
  {10.1146/annurev-conmatphys-033117-054235} (\bibinfo {year}
  {2017})\BibitemShut {NoStop}%
\bibitem [{\citenamefont {Paulose}\ \emph
  {et~al.}(2015{\natexlab{a}})\citenamefont {Paulose}, \citenamefont {Chen},\
  and\ \citenamefont {Vitelli}}]{Paulose2015}%
  \BibitemOpen
  \bibfield  {author} {\bibinfo {author} {\bibfnamefont {J.}~\bibnamefont
  {Paulose}}, \bibinfo {author} {\bibfnamefont {B.~G.~G.}\ \bibnamefont
  {Chen}},\ and\ \bibinfo {author} {\bibfnamefont {V.}~\bibnamefont
  {Vitelli}},\ }\bibfield  {title} {\bibinfo {title} {{Topological modes bound
  to dislocations in mechanical metamaterials}},\ }\bibfield  {journal}
  {\bibinfo  {journal} {Nature Physics}\ }\textbf {\bibinfo {volume} {11}},\
  \href {https://doi.org/10.1038/nphys3185} {10.1038/nphys3185} (\bibinfo
  {year} {2015}{\natexlab{a}})\BibitemShut {NoStop}%
\bibitem [{\citenamefont {Paulose}\ \emph
  {et~al.}(2015{\natexlab{b}})\citenamefont {Paulose}, \citenamefont
  {Meeussen},\ and\ \citenamefont {Vitelli}}]{Paulose2015a}%
  \BibitemOpen
  \bibfield  {author} {\bibinfo {author} {\bibfnamefont {J.}~\bibnamefont
  {Paulose}}, \bibinfo {author} {\bibfnamefont {A.~S.}\ \bibnamefont
  {Meeussen}},\ and\ \bibinfo {author} {\bibfnamefont {V.}~\bibnamefont
  {Vitelli}},\ }\bibfield  {title} {\bibinfo {title} {{Selective buckling via
  states of self-stress in topological metamaterials}},\ }\bibfield  {journal}
  {\bibinfo  {journal} {Proceedings of the National Academy of Sciences of the
  United States of America}\ }\href {https://doi.org/10.1073/pnas.1502939112}
  {10.1073/pnas.1502939112} (\bibinfo {year} {2015}{\natexlab{b}}),\ \Eprint
  {https://arxiv.org/abs/1502.03396} {arXiv:1502.03396} \BibitemShut {NoStop}%
\bibitem [{\citenamefont {Rocklin}\ \emph {et~al.}(2017)\citenamefont
  {Rocklin}, \citenamefont {Zhou}, \citenamefont {Sun},\ and\ \citenamefont
  {Mao}}]{Rocklin2017}%
  \BibitemOpen
  \bibfield  {author} {\bibinfo {author} {\bibfnamefont {D.~Z.}\ \bibnamefont
  {Rocklin}}, \bibinfo {author} {\bibfnamefont {S.}~\bibnamefont {Zhou}},
  \bibinfo {author} {\bibfnamefont {K.}~\bibnamefont {Sun}},\ and\ \bibinfo
  {author} {\bibfnamefont {X.}~\bibnamefont {Mao}},\ }\bibfield  {title}
  {\bibinfo {title} {{Transformable topological mechanical metamaterials}},\
  }\href {https://doi.org/10.1038/ncomms14201} {\bibfield  {journal} {\bibinfo
  {journal} {Nature Communications}\ }\textbf {\bibinfo {volume} {8}},\
  \bibinfo {pages} {1} (\bibinfo {year} {2017})},\ \Eprint
  {https://arxiv.org/abs/1510.06389} {arXiv:1510.06389} \BibitemShut {NoStop}%
\bibitem [{\citenamefont {Zhang}\ and\ \citenamefont {Mao}(2018)}]{Zhang2018}%
  \BibitemOpen
  \bibfield  {author} {\bibinfo {author} {\bibfnamefont {L.}~\bibnamefont
  {Zhang}}\ and\ \bibinfo {author} {\bibfnamefont {X.}~\bibnamefont {Mao}},\
  }\bibfield  {title} {\bibinfo {title} {{Fracturing of topological Maxwell
  lattices}},\ }\bibfield  {journal} {\bibinfo  {journal} {New Journal of
  Physics}\ }\textbf {\bibinfo {volume} {20}},\ \href
  {https://doi.org/10.1088/1367-2630/aac765} {10.1088/1367-2630/aac765}
  (\bibinfo {year} {2018})\BibitemShut {NoStop}%
\bibitem [{\citenamefont {Bi}\ \emph {et~al.}(2015)\citenamefont {Bi},
  \citenamefont {Lopez}, \citenamefont {Schwarz},\ and\ \citenamefont
  {Manning}}]{Bi2015}%
  \BibitemOpen
  \bibfield  {author} {\bibinfo {author} {\bibfnamefont {D.}~\bibnamefont
  {Bi}}, \bibinfo {author} {\bibfnamefont {J.~H.}\ \bibnamefont {Lopez}},
  \bibinfo {author} {\bibfnamefont {J.~M.}\ \bibnamefont {Schwarz}},\ and\
  \bibinfo {author} {\bibfnamefont {M.~L.}\ \bibnamefont {Manning}},\
  }\bibfield  {title} {\bibinfo {title} {{A density-independent rigidity
  transition in biological tissues}},\ }\bibfield  {journal} {\bibinfo
  {journal} {Nature Physics}\ }\href {https://doi.org/10.1038/nphys3471}
  {10.1038/nphys3471} (\bibinfo {year} {2015})\BibitemShut {NoStop}%
\bibitem [{\citenamefont {Bi}\ \emph {et~al.}(2016)\citenamefont {Bi},
  \citenamefont {Yang}, \citenamefont {Marchetti},\ and\ \citenamefont
  {Manning}}]{Bi2016}%
  \BibitemOpen
  \bibfield  {author} {\bibinfo {author} {\bibfnamefont {D.}~\bibnamefont
  {Bi}}, \bibinfo {author} {\bibfnamefont {X.}~\bibnamefont {Yang}}, \bibinfo
  {author} {\bibfnamefont {M.~C.}\ \bibnamefont {Marchetti}},\ and\ \bibinfo
  {author} {\bibfnamefont {M.~L.}\ \bibnamefont {Manning}},\ }\bibfield
  {title} {\bibinfo {title} {{Motility-driven glass and jamming transitions in
  biological tissues}},\ }\bibfield  {journal} {\bibinfo  {journal} {Physical
  Review X}\ }\href {https://doi.org/10.1103/PhysRevX.6.021011}
  {10.1103/PhysRevX.6.021011} (\bibinfo {year} {2016}),\ \Eprint
  {https://arxiv.org/abs/1509.06578} {arXiv:1509.06578} \BibitemShut {NoStop}%
\bibitem [{\citenamefont {Yan}\ and\ \citenamefont {Bi}(2019)}]{Yan2019}%
  \BibitemOpen
  \bibfield  {author} {\bibinfo {author} {\bibfnamefont {L.}~\bibnamefont
  {Yan}}\ and\ \bibinfo {author} {\bibfnamefont {D.}~\bibnamefont {Bi}},\
  }\bibfield  {title} {\bibinfo {title} {{Multicellular Rosettes Drive
  Fluid-solid Transition in Epithelial Tissues}},\ }\bibfield  {journal}
  {\bibinfo  {journal} {Physical Review X}\ }\href
  {https://doi.org/10.1103/PhysRevX.9.011029} {10.1103/PhysRevX.9.011029}
  (\bibinfo {year} {2019}),\ \Eprint {https://arxiv.org/abs/1806.04388}
  {arXiv:1806.04388} \BibitemShut {NoStop}%
\bibitem [{\citenamefont {Staple}\ \emph {et~al.}(2010)\citenamefont {Staple},
  \citenamefont {Farhadifar}, \citenamefont {Röper}, \citenamefont {Aigouy},
  \citenamefont {Eaton},\ and\ \citenamefont {Jülicher}}]{Staple2010}%
  \BibitemOpen
  \bibfield  {author} {\bibinfo {author} {\bibfnamefont {D.~B.}\ \bibnamefont
  {Staple}}, \bibinfo {author} {\bibfnamefont {R.}~\bibnamefont {Farhadifar}},
  \bibinfo {author} {\bibfnamefont {J.~C.}\ \bibnamefont {Röper}}, \bibinfo
  {author} {\bibfnamefont {B.}~\bibnamefont {Aigouy}}, \bibinfo {author}
  {\bibfnamefont {S.}~\bibnamefont {Eaton}},\ and\ \bibinfo {author}
  {\bibfnamefont {F.}~\bibnamefont {Jülicher}},\ }\bibfield  {title} {\bibinfo
  {title} {Mechanics and remodelling of cell packings in epithelia},\ }\href
  {https://doi.org/10.1140/epje/i2010-10677-0} {\bibfield  {journal} {\bibinfo
  {journal} {European Physical Journal E}\ }\textbf {\bibinfo {volume} {33}},\
  \bibinfo {pages} {117} (\bibinfo {year} {2010})}\BibitemShut {NoStop}%
\bibitem [{\citenamefont {Zhou}\ \emph {et~al.}(2018)\citenamefont {Zhou},
  \citenamefont {Zhang},\ and\ \citenamefont {Mao}}]{Zhou2018}%
  \BibitemOpen
  \bibfield  {author} {\bibinfo {author} {\bibfnamefont {D.}~\bibnamefont
  {Zhou}}, \bibinfo {author} {\bibfnamefont {L.}~\bibnamefont {Zhang}},\ and\
  \bibinfo {author} {\bibfnamefont {X.}~\bibnamefont {Mao}},\ }\bibfield
  {title} {\bibinfo {title} {{Topological Edge Floppy Modes in Disordered Fiber
  Networks}},\ }\bibfield  {journal} {\bibinfo  {journal} {Physical Review
  Letters}\ }\textbf {\bibinfo {volume} {120}},\ \href
  {https://doi.org/10.1103/PhysRevLett.120.068003}
  {10.1103/PhysRevLett.120.068003} (\bibinfo {year} {2018}),\ \Eprint
  {https://arxiv.org/abs/1708.03935} {arXiv:1708.03935} \BibitemShut {NoStop}%
\bibitem [{\citenamefont {Chiou}\ \emph {et~al.}(2012)\citenamefont {Chiou},
  \citenamefont {Hufnagel},\ and\ \citenamefont {Shraiman}}]{Chiou2012}%
  \BibitemOpen
  \bibfield  {author} {\bibinfo {author} {\bibfnamefont {K.~K.}\ \bibnamefont
  {Chiou}}, \bibinfo {author} {\bibfnamefont {L.}~\bibnamefont {Hufnagel}},\
  and\ \bibinfo {author} {\bibfnamefont {B.~I.}\ \bibnamefont {Shraiman}},\
  }\bibfield  {title} {\bibinfo {title} {{Mechanical stress inference for two
  dimensional cell arrays}},\ }\bibfield  {journal} {\bibinfo  {journal} {PLoS
  Computational Biology}\ }\href {https://doi.org/10.1371/journal.pcbi.1002512}
  {10.1371/journal.pcbi.1002512} (\bibinfo {year} {2012})\BibitemShut {NoStop}%
\bibitem [{\citenamefont {Noll}\ \emph {et~al.}(2017)\citenamefont {Noll},
  \citenamefont {Mani}, \citenamefont {Heemskerk}, \citenamefont {Streichan},\
  and\ \citenamefont {Shraiman}}]{Noll2017}%
  \BibitemOpen
  \bibfield  {author} {\bibinfo {author} {\bibfnamefont {N.}~\bibnamefont
  {Noll}}, \bibinfo {author} {\bibfnamefont {M.}~\bibnamefont {Mani}}, \bibinfo
  {author} {\bibfnamefont {I.}~\bibnamefont {Heemskerk}}, \bibinfo {author}
  {\bibfnamefont {S.~J.}\ \bibnamefont {Streichan}},\ and\ \bibinfo {author}
  {\bibfnamefont {B.~I.}\ \bibnamefont {Shraiman}},\ }\bibfield  {title}
  {\bibinfo {title} {{Active tension network model suggests an exotic
  mechanical state realized in epithelial tissues}},\ }\bibfield  {journal}
  {\bibinfo  {journal} {Nature Physics}\ }\href
  {https://doi.org/10.1038/nphys4219} {10.1038/nphys4219} (\bibinfo {year}
  {2017}),\ \Eprint {https://arxiv.org/abs/1508.00623} {arXiv:1508.00623}
  \BibitemShut {NoStop}%
\bibitem [{\citenamefont {Hayes}\ and\ \citenamefont
  {Solon}(2017)}]{Hayes2017}%
  \BibitemOpen
  \bibfield  {author} {\bibinfo {author} {\bibfnamefont {P.}~\bibnamefont
  {Hayes}}\ and\ \bibinfo {author} {\bibfnamefont {J.}~\bibnamefont {Solon}},\
  }\href {https://doi.org/10.1016/j.mod.2016.12.005} {\bibinfo {title}
  {Drosophila dorsal closure: An orchestra of forces to zip shut the embryo}}
  (\bibinfo {year} {2017})\BibitemShut {NoStop}%
\bibitem [{\citenamefont {Kiehart}\ \emph {et~al.}(2017)\citenamefont
  {Kiehart}, \citenamefont {Crawford}, \citenamefont {Aristotelous},
  \citenamefont {Venakides},\ and\ \citenamefont {Edwards}}]{Kiehart2017}%
  \BibitemOpen
  \bibfield  {author} {\bibinfo {author} {\bibfnamefont {D.~P.}\ \bibnamefont
  {Kiehart}}, \bibinfo {author} {\bibfnamefont {J.~M.}\ \bibnamefont
  {Crawford}}, \bibinfo {author} {\bibfnamefont {A.}~\bibnamefont
  {Aristotelous}}, \bibinfo {author} {\bibfnamefont {S.}~\bibnamefont
  {Venakides}},\ and\ \bibinfo {author} {\bibfnamefont {G.~S.}\ \bibnamefont
  {Edwards}},\ }\bibfield  {title} {\bibinfo {title} {Cell sheet morphogenesis:
  Dorsal closure in drosophila melanogaster as a model system},\ }\href
  {https://doi.org/10.1146/annurev-cellbio-111315-125357} {\bibfield  {journal}
  {\bibinfo  {journal} {Annual Review of Cell and Developmental Biology}\
  }\textbf {\bibinfo {volume} {33}},\ \bibinfo {pages} {169} (\bibinfo {year}
  {2017})}\BibitemShut {NoStop}%
\bibitem [{\citenamefont {Ninov}\ \emph {et~al.}(2007)\citenamefont {Ninov},
  \citenamefont {Chiarelli},\ and\ \citenamefont {Martín-Blanco}}]{Ninov2007}%
  \BibitemOpen
  \bibfield  {author} {\bibinfo {author} {\bibfnamefont {N.}~\bibnamefont
  {Ninov}}, \bibinfo {author} {\bibfnamefont {D.~A.}\ \bibnamefont
  {Chiarelli}},\ and\ \bibinfo {author} {\bibfnamefont {E.}~\bibnamefont
  {Martín-Blanco}},\ }\bibfield  {title} {\bibinfo {title} {Extrinsic and
  intrinsic mechanisms directing epithelial cell sheet replacement during
  drosophila metamorphosis},\ }\href {https://doi.org/10.1242/dev.02728}
  {\bibfield  {journal} {\bibinfo  {journal} {Development}\ }\textbf {\bibinfo
  {volume} {134}},\ \bibinfo {pages} {367} (\bibinfo {year}
  {2007})}\BibitemShut {NoStop}%
\bibitem [{\citenamefont {Ainslie}\ \emph {et~al.}(2020)\citenamefont
  {Ainslie}, \citenamefont {Davis}, \citenamefont {Williamson}, \citenamefont
  {Ferreira}, \citenamefont {Torres-Sánchez}, \citenamefont {Hoppe},
  \citenamefont {Mangione}, \citenamefont {Smith}, \citenamefont
  {Martin-Blanco}, \citenamefont {Salbreux},\ and\ \citenamefont
  {Tapon}}]{Ainslie2020}%
  \BibitemOpen
  \bibfield  {author} {\bibinfo {author} {\bibfnamefont {A.~P.}\ \bibnamefont
  {Ainslie}}, \bibinfo {author} {\bibfnamefont {J.~R.}\ \bibnamefont {Davis}},
  \bibinfo {author} {\bibfnamefont {J.~J.}\ \bibnamefont {Williamson}},
  \bibinfo {author} {\bibfnamefont {A.}~\bibnamefont {Ferreira}}, \bibinfo
  {author} {\bibfnamefont {A.}~\bibnamefont {Torres-Sánchez}}, \bibinfo
  {author} {\bibfnamefont {A.}~\bibnamefont {Hoppe}}, \bibinfo {author}
  {\bibfnamefont {F.}~\bibnamefont {Mangione}}, \bibinfo {author}
  {\bibfnamefont {M.~B.}\ \bibnamefont {Smith}}, \bibinfo {author}
  {\bibfnamefont {E.}~\bibnamefont {Martin-Blanco}}, \bibinfo {author}
  {\bibfnamefont {G.}~\bibnamefont {Salbreux}},\ and\ \bibinfo {author}
  {\bibfnamefont {N.}~\bibnamefont {Tapon}},\ }\href
  {https://doi.org/10.1101/2020.11.10.376129} {\bibinfo {title} {Ecm remodeling
  and spatial cell cycle coordination determine tissue growth kinetics}}
  (\bibinfo {year} {2020})\BibitemShut {NoStop}%
\bibitem [{\citenamefont {Bruce}\ and\ \citenamefont
  {Heisenberg}(2020)}]{Bruce2020}%
  \BibitemOpen
  \bibfield  {author} {\bibinfo {author} {\bibfnamefont {A.~E.}\ \bibnamefont
  {Bruce}}\ and\ \bibinfo {author} {\bibfnamefont {C.~P.}\ \bibnamefont
  {Heisenberg}},\ }\href {https://doi.org/10.1016/bs.ctdb.2019.07.001}
  {\bibinfo {title} {Mechanisms of zebrafish epiboly: A current view}}
  (\bibinfo {year} {2020})\BibitemShut {NoStop}%
\bibitem [{\citenamefont {Begnaud}\ \emph {et~al.}(2016)\citenamefont
  {Begnaud}, \citenamefont {Chen}, \citenamefont {Delacour}, \citenamefont
  {Mège},\ and\ \citenamefont {Ladoux}}]{Begnaud2016}%
  \BibitemOpen
  \bibfield  {author} {\bibinfo {author} {\bibfnamefont {S.}~\bibnamefont
  {Begnaud}}, \bibinfo {author} {\bibfnamefont {T.}~\bibnamefont {Chen}},
  \bibinfo {author} {\bibfnamefont {D.}~\bibnamefont {Delacour}}, \bibinfo
  {author} {\bibfnamefont {R.~M.}\ \bibnamefont {Mège}},\ and\ \bibinfo
  {author} {\bibfnamefont {B.}~\bibnamefont {Ladoux}},\ }\href
  {https://doi.org/10.1016/j.ceb.2016.04.006} {\bibinfo {title} {Mechanics of
  epithelial tissues during gap closure}} (\bibinfo {year} {2016})\BibitemShut
  {NoStop}%
\bibitem [{\citenamefont {Hakim}\ and\ \citenamefont
  {Silberzan}(2017)}]{Hakim2017}%
  \BibitemOpen
  \bibfield  {author} {\bibinfo {author} {\bibfnamefont {V.}~\bibnamefont
  {Hakim}}\ and\ \bibinfo {author} {\bibfnamefont {P.}~\bibnamefont
  {Silberzan}},\ }\href {https://doi.org/10.1088/1361-6633/aa65ef} {\bibinfo
  {title} {Collective cell migration: A physics perspective}} (\bibinfo {year}
  {2017})\BibitemShut {NoStop}%
\bibitem [{\citenamefont {Honda}(1983)}]{Honda1983}%
  \BibitemOpen
  \bibfield  {author} {\bibinfo {author} {\bibfnamefont {H.}~\bibnamefont
  {Honda}},\ }\bibfield  {title} {\bibinfo {title} {Geometrical models for
  cells in tissues},\ }in\ \href@noop {} {\emph {\bibinfo {booktitle}
  {International review of cytology}}},\ Vol.~\bibinfo {volume} {81}\ (\bibinfo
   {publisher} {Elsevier},\ \bibinfo {year} {1983})\ pp.\ \bibinfo {pages}
  {191--248}\BibitemShut {NoStop}%
\bibitem [{\citenamefont {Farhadifar}\ \emph {et~al.}(2007)\citenamefont
  {Farhadifar}, \citenamefont {R{\"o}per}, \citenamefont {Aigouy},
  \citenamefont {Eaton},\ and\ \citenamefont {J{\"u}licher}}]{Farhadifar2007}%
  \BibitemOpen
  \bibfield  {author} {\bibinfo {author} {\bibfnamefont {R.}~\bibnamefont
  {Farhadifar}}, \bibinfo {author} {\bibfnamefont {J.-C.}\ \bibnamefont
  {R{\"o}per}}, \bibinfo {author} {\bibfnamefont {B.}~\bibnamefont {Aigouy}},
  \bibinfo {author} {\bibfnamefont {S.}~\bibnamefont {Eaton}},\ and\ \bibinfo
  {author} {\bibfnamefont {F.}~\bibnamefont {J{\"u}licher}},\ }\bibfield
  {title} {\bibinfo {title} {The influence of cell mechanics, cell-cell
  interactions, and proliferation on epithelial packing},\ }\href@noop {}
  {\bibfield  {journal} {\bibinfo  {journal} {Current Biology}\ }\textbf
  {\bibinfo {volume} {17}},\ \bibinfo {pages} {2095} (\bibinfo {year}
  {2007})}\BibitemShut {NoStop}%
\bibitem [{\citenamefont {Fletcher}\ \emph {et~al.}(2014)\citenamefont
  {Fletcher}, \citenamefont {Osterfield}, \citenamefont {Baker},\ and\
  \citenamefont {Shvartsman}}]{Fletcher2014}%
  \BibitemOpen
  \bibfield  {author} {\bibinfo {author} {\bibfnamefont {A.~G.}\ \bibnamefont
  {Fletcher}}, \bibinfo {author} {\bibfnamefont {M.}~\bibnamefont
  {Osterfield}}, \bibinfo {author} {\bibfnamefont {R.~E.}\ \bibnamefont
  {Baker}},\ and\ \bibinfo {author} {\bibfnamefont {S.~Y.}\ \bibnamefont
  {Shvartsman}},\ }\href {https://doi.org/10.1016/j.bpj.2013.11.4498} {\bibinfo
  {title} {{Vertex models of epithelial morphogenesis}}} (\bibinfo {year}
  {2014})\BibitemShut {NoStop}%
\bibitem [{\citenamefont {Yang}\ \emph {et~al.}(2017)\citenamefont {Yang},
  \citenamefont {Bi}, \citenamefont {Czajkowski}, \citenamefont {Merkel},
  \citenamefont {Manning},\ and\ \citenamefont {Marchetti}}]{Yang2017}%
  \BibitemOpen
  \bibfield  {author} {\bibinfo {author} {\bibfnamefont {X.}~\bibnamefont
  {Yang}}, \bibinfo {author} {\bibfnamefont {D.}~\bibnamefont {Bi}}, \bibinfo
  {author} {\bibfnamefont {M.}~\bibnamefont {Czajkowski}}, \bibinfo {author}
  {\bibfnamefont {M.}~\bibnamefont {Merkel}}, \bibinfo {author} {\bibfnamefont
  {M.~L.}\ \bibnamefont {Manning}},\ and\ \bibinfo {author} {\bibfnamefont
  {M.~C.}\ \bibnamefont {Marchetti}},\ }\bibfield  {title} {\bibinfo {title}
  {{Correlating cell shape and cellular stress in motile confluent tissues}},\
  }\bibfield  {journal} {\bibinfo  {journal} {Proceedings of the National
  Academy of Sciences of the United States of America}\ }\href
  {https://doi.org/10.1073/pnas.1705921114} {10.1073/pnas.1705921114} (\bibinfo
  {year} {2017}),\ \Eprint {https://arxiv.org/abs/1704.05951}
  {arXiv:1704.05951} \BibitemShut {NoStop}%
\bibitem [{\citenamefont {Merkel}\ and\ \citenamefont
  {Manning}(2018)}]{Merkel2018}%
  \BibitemOpen
  \bibfield  {author} {\bibinfo {author} {\bibfnamefont {M.}~\bibnamefont
  {Merkel}}\ and\ \bibinfo {author} {\bibfnamefont {M.~L.}\ \bibnamefont
  {Manning}},\ }\bibfield  {title} {\bibinfo {title} {{A geometrically
  controlled rigidity transition in a model for confluent 3D tissues}},\
  }\bibfield  {journal} {\bibinfo  {journal} {New Journal of Physics}\ }\href
  {https://doi.org/10.1088/1367-2630/aaaa13} {10.1088/1367-2630/aaaa13}
  (\bibinfo {year} {2018}),\ \Eprint {https://arxiv.org/abs/1706.02656}
  {arXiv:1706.02656} \BibitemShut {NoStop}%
\bibitem [{\citenamefont {Hufnagel}\ \emph {et~al.}(2007)\citenamefont
  {Hufnagel}, \citenamefont {Teleman}, \citenamefont {Rouault}, \citenamefont
  {Cohen},\ and\ \citenamefont {Shraiman}}]{Hufnagel2007}%
  \BibitemOpen
  \bibfield  {author} {\bibinfo {author} {\bibfnamefont {L.}~\bibnamefont
  {Hufnagel}}, \bibinfo {author} {\bibfnamefont {A.~A.}\ \bibnamefont
  {Teleman}}, \bibinfo {author} {\bibfnamefont {H.}~\bibnamefont {Rouault}},
  \bibinfo {author} {\bibfnamefont {S.~M.}\ \bibnamefont {Cohen}},\ and\
  \bibinfo {author} {\bibfnamefont {B.~I.}\ \bibnamefont {Shraiman}},\
  }\bibfield  {title} {\bibinfo {title} {On the mechanism of wing size
  determination in fly development},\ }\href
  {https://doi.org/10.1073/pnas.0607134104} {\bibfield  {journal} {\bibinfo
  {journal} {Proceedings of the National Academy of Sciences of the United
  States of America}\ }\textbf {\bibinfo {volume} {104}},\ \bibinfo {pages}
  {3835} (\bibinfo {year} {2007})}\BibitemShut {NoStop}%
\bibitem [{\citenamefont {Salbreux}\ \emph {et~al.}(2012)\citenamefont
  {Salbreux}, \citenamefont {Barthel}, \citenamefont {Raymond},\ and\
  \citenamefont {Lubensky}}]{Salbreux2012}%
  \BibitemOpen
  \bibfield  {author} {\bibinfo {author} {\bibfnamefont {G.}~\bibnamefont
  {Salbreux}}, \bibinfo {author} {\bibfnamefont {L.~K.}\ \bibnamefont
  {Barthel}}, \bibinfo {author} {\bibfnamefont {P.~A.}\ \bibnamefont
  {Raymond}},\ and\ \bibinfo {author} {\bibfnamefont {D.~K.}\ \bibnamefont
  {Lubensky}},\ }\bibfield  {title} {\bibinfo {title} {Coupling mechanical
  deformations and planar cell polarity to create regular patterns in the
  zebrafish retina},\ }\href {https://doi.org/10.1371/journal.pcbi.1002618}
  {\bibfield  {journal} {\bibinfo  {journal} {PLoS Computational Biology}\
  }\textbf {\bibinfo {volume} {8}},\ \bibinfo {pages} {1002618} (\bibinfo
  {year} {2012})}\BibitemShut {NoStop}%
\bibitem [{\citenamefont {Spencer}\ \emph {et~al.}(2017)\citenamefont
  {Spencer}, \citenamefont {Jabeen},\ and\ \citenamefont
  {Lubensky}}]{Spencer2017}%
  \BibitemOpen
  \bibfield  {author} {\bibinfo {author} {\bibfnamefont {M.~A.}\ \bibnamefont
  {Spencer}}, \bibinfo {author} {\bibfnamefont {Z.}~\bibnamefont {Jabeen}},\
  and\ \bibinfo {author} {\bibfnamefont {D.~K.}\ \bibnamefont {Lubensky}},\
  }\bibfield  {title} {\bibinfo {title} {Vertex stability and topological
  transitions in vertex models of foams and epithelia},\ }\href
  {https://doi.org/10.1140/epje/i2017-11489-4} {\bibfield  {journal} {\bibinfo
  {journal} {European Physical Journal E}\ }\textbf {\bibinfo {volume} {40}},\
  \bibinfo {pages} {1} (\bibinfo {year} {2017})}\BibitemShut {NoStop}%
\bibitem [{\citenamefont {Stenull}\ and\ \citenamefont
  {Lubensky}(2019)}]{Stenull2019}%
  \BibitemOpen
  \bibfield  {author} {\bibinfo {author} {\bibfnamefont {O.}~\bibnamefont
  {Stenull}}\ and\ \bibinfo {author} {\bibfnamefont {T.~C.}\ \bibnamefont
  {Lubensky}},\ }\bibfield  {title} {\bibinfo {title} {Signatures of
  topological phonons in superisostatic lattices},\ }\bibfield  {journal}
  {\bibinfo  {journal} {Physical Review Letters}\ }\textbf {\bibinfo {volume}
  {122}},\ \href {https://doi.org/10.1103/PhysRevLett.122.248002}
  {10.1103/PhysRevLett.122.248002} (\bibinfo {year} {2019})\BibitemShut
  {NoStop}%
\bibitem [{\citenamefont {Sun}\ and\ \citenamefont {Mao}(2020)}]{Sun2020}%
  \BibitemOpen
  \bibfield  {author} {\bibinfo {author} {\bibfnamefont {K.}~\bibnamefont
  {Sun}}\ and\ \bibinfo {author} {\bibfnamefont {X.}~\bibnamefont {Mao}},\
  }\bibfield  {title} {\bibinfo {title} {Continuum theory for topological edge
  soft modes},\ }\bibfield  {journal} {\bibinfo  {journal} {Physical Review
  Letters}\ }\textbf {\bibinfo {volume} {124}},\ \href
  {https://doi.org/10.1103/PhysRevLett.124.207601}
  {10.1103/PhysRevLett.124.207601} (\bibinfo {year} {2020})\BibitemShut
  {NoStop}%
\bibitem [{\citenamefont {Saremi}\ and\ \citenamefont
  {Rocklin}(2020)}]{Saremi2020}%
  \BibitemOpen
  \bibfield  {author} {\bibinfo {author} {\bibfnamefont {A.}~\bibnamefont
  {Saremi}}\ and\ \bibinfo {author} {\bibfnamefont {Z.}~\bibnamefont
  {Rocklin}},\ }\bibfield  {title} {\bibinfo {title} {Topological elasticity of
  flexible structures},\ }\bibfield  {journal} {\bibinfo  {journal} {Physical
  Review X}\ }\textbf {\bibinfo {volume} {10}},\ \href
  {https://doi.org/10.1103/PhysRevX.10.011052} {10.1103/PhysRevX.10.011052}
  (\bibinfo {year} {2020})\BibitemShut {NoStop}%
\bibitem [{\citenamefont {Calladine}(1978)}]{Calladine1978}%
  \BibitemOpen
  \bibfield  {author} {\bibinfo {author} {\bibfnamefont {C.~R.}\ \bibnamefont
  {Calladine}},\ }\bibfield  {title} {\bibinfo {title} {{Buckminster Fuller's
  "Tensegrity" structures and Clerk Maxwell's rules for the construction of
  stiff frames}},\ }\bibfield  {journal} {\bibinfo  {journal} {International
  Journal of Solids and Structures}\ }\href
  {https://doi.org/10.1016/0020-7683(78)90052-5} {10.1016/0020-7683(78)90052-5}
  (\bibinfo {year} {1978})\BibitemShut {NoStop}%
\bibitem [{\citenamefont {Zhou}\ \emph {et~al.}(2019)\citenamefont {Zhou},
  \citenamefont {Zhang},\ and\ \citenamefont {Mao}}]{Zhou2019}%
  \BibitemOpen
  \bibfield  {author} {\bibinfo {author} {\bibfnamefont {D.}~\bibnamefont
  {Zhou}}, \bibinfo {author} {\bibfnamefont {L.}~\bibnamefont {Zhang}},\ and\
  \bibinfo {author} {\bibfnamefont {X.}~\bibnamefont {Mao}},\ }\bibfield
  {title} {\bibinfo {title} {{Topological Boundary Floppy Modes in
  Quasicrystals}},\ }\bibfield  {journal} {\bibinfo  {journal} {Physical Review
  X}\ }\textbf {\bibinfo {volume} {9}},\ \href
  {https://doi.org/10.1103/PhysRevX.9.021054} {10.1103/PhysRevX.9.021054}
  (\bibinfo {year} {2019})\BibitemShut {NoStop}%
\bibitem [{\citenamefont {Rocklin}(2017)}]{rocklin2017directional}%
  \BibitemOpen
  \bibfield  {author} {\bibinfo {author} {\bibfnamefont {D.~Z.}\ \bibnamefont
  {Rocklin}},\ }\bibfield  {title} {\bibinfo {title} {Directional mechanical
  response in the bulk of topological metamaterials},\ }\href@noop {}
  {\bibfield  {journal} {\bibinfo  {journal} {New Journal of Physics}\ }\textbf
  {\bibinfo {volume} {19}},\ \bibinfo {pages} {065004} (\bibinfo {year}
  {2017})}\BibitemShut {NoStop}%
\end{thebibliography}%

\end{document}